\documentclass[twocolumn,preprintnumbers,amsmath,amssymb,superscriptaddress,nofootinbib,longbibliography, prb]{revtex4-2}

\usepackage{graphicx}
\usepackage{bm}
\usepackage{dsfont}
\usepackage[usenames,dvipsnames]{xcolor}
\usepackage{pstricks}
\usepackage[tight]{subfigure}
\usepackage{verbatim}
\usepackage{units}
\usepackage{multirow}
\usepackage{mathrsfs}
\usepackage{leftidx}
\usepackage{xspace}
\usepackage{diffcoeff}  
\usepackage{bbm}
\usepackage{amsfonts,amssymb,amsmath}
\usepackage{mathtools}

\usepackage{array} 
\usepackage{tabularx}
\usepackage[normalem]{ulem}
\usepackage{bbold}
\usepackage{pifont}
\usepackage{hyperref}
\hypersetup{colorlinks=true,linktoc=all,linkcolor=blue,breaklinks=true,citecolor=blue,urlcolor=blue}

\usepackage[utf8]{inputenc}
\usepackage[english]{babel}

\definecolor{darkgreen}{RGB}{6, 153, 38}

\addto\captionsenglish{}

\AtBeginDocument{%
    \newwrite\bibnotes
    \def\bibnotesext{Notes.bib}
    \immediate\openout\bibnotes=\jobname\bibnotesext
    \immediate\write\bibnotes{@CONTROL{REVTEX42Control}}
    \immediate\write\bibnotes{@CONTROL{%
    apsrev42Control,author="08",editor="1",pages="1",title="0",year="1"}}
     \if@filesw
     \immediate\write\@auxout{\string\citation{apsrev42Control}}%
    \fi
}%

\newcommand{\beq}{\begin{equation}}
\newcommand{\eeq}{\end{equation}}

\DeclarePairedDelimiterX\braket[2]{\langle}{\rangle}{#1\,\delimsize\vert\,\mathopen{}#2}
\DeclarePairedDelimiterX\ketbra[2]{\lvert}{\rvert}{#1\,\delimsize\rangle\mathopen{}\delimsize\langle\,\mathopen{}#2}

\DeclarePairedDelimiterX\Braket[2]{(}{)}{#1\,\delimsize\vert\,\mathopen{}#2}
\DeclarePairedDelimiterX\Ketbra[2]{\lvert}{\rvert}{#1\,\delimsize)\mathopen{}\delimsize(\,\mathopen{}#2}

\newcommand{\kB}{k_\text{B}}
\newcommand{\kBT}{k_\text{B}T}

\begin{document}

\title[Quantum thermocouples]{Quantum thermocouples: nonlocal conversion and control of heat in nanostructures}


\author{Jos\'e Balduque}
\affiliation{Departamento de F\'isica Te\'orica de la Materia Condensada, Universidad Aut\'onoma de Madrid, 28049 Madrid, Spain\looseness=-1}
\affiliation{Condensed Matter Physics Center (IFIMAC), Universidad Aut\'onoma de Madrid, 28049 Madrid, Spain\looseness=-1}
\author{Rafael S\'anchez}
\affiliation{Departamento de F\'isica Te\'orica de la Materia Condensada, Universidad Aut\'onoma de Madrid, 28049 Madrid, Spain\looseness=-1}
\affiliation{Condensed Matter Physics Center (IFIMAC), Universidad Aut\'onoma de Madrid, 28049 Madrid, Spain\looseness=-1}
\affiliation{Instituto Nicol\'as Cabrera (INC), Universidad Aut\'onoma de Madrid, 28049 Madrid, Spain\looseness=-1}

\begin{abstract}
Nanoscale conductors are interesting for thermoelectrics because of their particular spectral features connecting separated heat and particle currents. Multiterminal devices in the quantum regime benefit from phase-coherent phenomena, which turns the thermoelectric effect nonlocal, and from tunable single-particle interactions. This way one can define quantum thermocouples which convert an injected heat current into useful power in an isothermal conductor, or work as refrigerators. Additionally, efficient heat management devices can be defined.
We review recent theoretical and experimental progress in the research of multiterminal thermal and thermoelectric quantum transport leading to proposals of autonomous quantum heat engines and thermal devices.
\end{abstract}




\maketitle

\section{Introduction}\label{sec:intro}
\begin{figure*}[t]
\centering
\includegraphics[width=\textwidth]{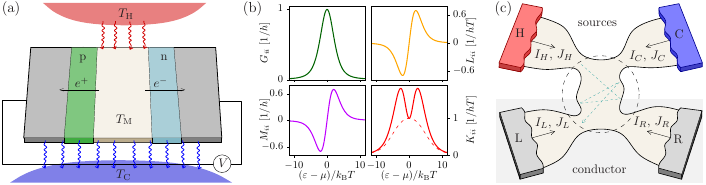}
\caption{(a) Macroscopic thermocouple. The central region of a conductor otherwise at a temperature $T_C$ is in contact with a heat source at temperature $T_H$ for which reason it reaches a local equilibrium at temperature $T_M$. Electron-hole excitations generate a thermovoltage when separated by p and n junctions. (b) Onsager coefficients $O=O_{ii}$ of a two-terminal resonant-tunneling barrier when sweeping the resonance energy $\varepsilon$ across the electrochemical potential $\mu$, with rates $\Gamma_{1}=\Gamma_{2}=\kBT$, cf. Eq.~\eqref{eq:lorentz}. The thermal conductance in the absence of particle current, $\tilde K=K-LM/G$ is plotted as a red dashed line. (c) Quantum thermocouple. An isothermal conductor is formed by terminals L and R. It is connected to heat sources (labeled H and C) via a mesoscopic region (dashed circle) where the electron phase coherence is preserved. Nonlocal transport $I_L=-I_R$ is a consequence of the (interference) properties of scattering in the mesoscopic region. }\label{fig:macrothcpl}
\end{figure*}

The thermocouple is the minimal thermoelectric device able to use the Seebeck effect~\cite{ziman_principles_1972} to transduce a heat flow into an electric response measurable in an isothermal conductor. It is useful for the autonomous conversion of environmental (or waste) heat into useful power~\cite{Snyder2008Feb,Bell2008Sep}, or as a temperature sensor when coupled to a voltmeter. The basic principle of a macroscopic thermocouple is sketched in Fig.~\ref{fig:macrothcpl}(a). Two junctions separate the conductor, at a base temperature $T_C$, from a region that is in contact with a heat source at temperature $T_H$. If the two junctions have opposite thermoelectric response (e.g., one being n-type and the other p-type), the electron-hole excitations, thermalized at a temperature $T_M$ in the central region, are spatially separated to opposite contacts. 
This can then either generate a current or be measured as a voltage difference which depends on the difference of the linear Seebeck coefficient of each junction: $S_p-S_n$. The reverse process is also possible: driving a current through the conductor can be used to extract heat from the central region, in which case one has a Peltier refrigerator. 

Reducing the size of the thermocouple to nanometer scales~\cite{heremans_when_2013} comes with a number of key differences. 
The first to be noted was that the reduced dimensionality provides the required spectral properties for efficient thermoelectric response~\cite{hicks_effect_1993,hicks:1993}, which can also be easily tuned with gate voltages, as illustrated in Fig.~\ref{fig:macrothcpl}(b) for a resonant-tunneling barrier.
At this scale, it becomes difficult to maintain large temperature differences. However, one benefits from the system response easily becoming nonlinear. 
From the point of view of the physical processes, the size of the region in contact with the heat source can be of the order of or smaller than the carrier thermalization length. This has two possible consequences: the thermoelectric response may become nonlocal~\cite{bjornreview,benenti:2017} (not determined by the series response of the junctions) or determined by nonthermal states~\cite{whitney_quantum_2019}, as represented in Fig.~\ref{fig:macrothcpl}(c). Finally, at low enough temperatures quantum properties arising from electrons propagating as waves in conductors smaller than the phase coherence length start to contribute, leading to interference effects~\cite{heikkila_book,moskalets-book}. Also, single particle interactions (electron-electron or electron-photon) can be relevant~\cite{bruus_book}. This is the regime where we focus our interest in this review. Properties from quantum materials (e.g., superconductors or edge states in quantum Hall regimes, spin-orbit couplings) can also be exploited~\cite{benenti:2017,arrachea_thermoelectric_2025,bauer_spin_2012,yang_role_2023}.   

While the electronic transport through nanostructures has been intensively investigated during the last decades both experimental~\cite{ihn_semiconductor_2009} and theoretically~\cite{nazarov_quantum_2009}, progress in the control of thermal and thermoelectric flows was somehow delayed, mainly due to the difficulty in controlling and measuring minute temperature differences in an experiment and in eliminating phonon drag contributions. 
This has however changed in the last years~\cite{giazotto:2006,dubi_thermoelectric_2009,pekola_colloquium_2021,majidi_heat_2024}. The interest is not restricted to the possibility to generate useful thermodynamic outputs but also to use the thermoelectric response as sensitive probes for transport mechanisms, in particular involving the difficult to measure energy flow. 
Controlled thermoelectric measurements have been achieved in 2DEGs~\cite{fletcher_oscillations_1995,chickering_hot_2009,schmidt_thermopower_2012,jain_heating_2025}, quantum point contacts~\cite{molenkamp:1990,molenkamp:1992,vanHouten:1992,brun2019}, quantum dots~\cite{staring_coulomb_1993,dzurak_observation_1993,dzurak_thermoelectric_1997,scheibner_thermopower_2005,scheibner_sequential_2007,svensson_lineshape_2012,svensson_nonlinear_2013,thierschmann_diffusion_2013,josefsson_quantum_2018,prete:2019,dutta_direct_2019,kleeorin:2019,asgari_quantumdot_2021}, single-electron transistors~\cite{erdman:2019}, nanowires~\cite{appleyard_direction_2000,hochbaum_enhanced_2008,curtin_highly_2012,stranz_high_2013,karg_measurement_2013,roddaro_giant_2013,brovman_electric_2016,chen:2018,fast:2020,gachter_spatially_2020,zolotavin:2017,fast_geometric_2024}, scanning probes~\cite{Park2013,harzheim:2018}, molecular junctions~\cite{rinconGarcia_thermopower_2016,cui_perspective_2017,miao_influence_2018,volosheniuk_enjancing_2025,volosheniuk_single_2025}, superconducting tunnel junctions~\cite{kolenda:2016,kolenda:2017,germanese_bipolar_2022} or magnetic wires~\cite{puttock_local_2022}, also in cold atom gases~\cite{brantut_thermoelectric_2013}.
In particular, three-terminal quantum dot arrangements in thermocouple geometries have been realized using electron-electron interactions~\cite{holger,roche:2015,hartmann:2015}, hot cavities~\cite{jaliel:2019}, electron-phonon~\cite{dorsch:2020,dorsch_characterization_2021} or electron-photon interactions~\cite{haldar_microwave_2024,stanisavljevic_efficient_2024} to couple the conductor to the heat source. Multiterminal quantum Hall edge states~\cite{granger_observation_2009,nam_thermoelectric_2013,huynh_chiral_2025}, (local) Joule heating in graphene~\cite{mitra_anomalous_2021} and normal-superconductor interfaces~\cite{Tan2021} have also been considered. 
Interestingly, similar responses have been measured when coupling the conductor to nonequilibrium fluctuations~\cite{khrapai_doubledot_2006,yamamoto_negative_2006,khrapai_counterflow_2007,khrapai_nonequilibrium_2008,hashisaka_bolometric_2008,harbusch_phonon_2010,laroche_positive_2011,laroche_1d1d_2014,bischoff:2015,li_negative_2016,keller:2016,takada_heatdriven_2021,makaju_nonreciprocal_2024,fu_nonreciprocal_2025}, which can be related to energy quanta emitted by relaxing electrons~\cite{taubert_relaxation_2011}.

This progress has been accompanied by theoretical works proposing the multiterminal thermoelectric response to take advantage of the separation of thermal and particle currents~\cite{hotspots,mazza:2015,panu_heatcharge_2024} or to access properties not available in the electrical conductance, for instance nonabelian statistics of carriers~\cite{yang_thermopower_2009}, odd-frequency superconducting correlations~\cite{hwang:2018}, crossed-Andreev reflection~\cite{sanchez_cooling_2018,Hussein2019,Kirsanov2019}, correlations between edge states~\cite{braggio_nonlocal_2024,yamazaki_efficient_2025} or details of interfaces between superconductors and topological materials~\cite{Blasi2020a,blasi_nonlocal_2020b,mateos_nonlocal_2024}. At the same time, questions related to quantum thermodynamics~\cite{pekola_toward_2015}, e.g. the efficiency of quantum heat engines~\cite{whitney_most_2014,whitney_finding_2015}, or the role of fluctuations~\cite{sanchez_detection_2012,ptaszynski_coherence_2018,kheradsoud:2019,balduque_coherent_2024,tesser_out_2024,acciai_constraints_2024} and of information~\cite{hotspots,strasberg:2013,horowitz:2014prx,kutvonen:2016,sanchez:2019,whitney_illusory_2023,schaller_how_2024,monsel_autonomous_2025} have found in the thermoelectric effect a suitable platform to be addressed. 

Particularly relevant to our discussion in this review is the configuration where a quantum system is coupled to two electronic terminals at the same temperature, with a third terminal (or bath) injecting only heat. Under the appropriate broken symmetries, the injected heat is rectified into a particle current. Initial proposals considered electron-phonon coupling~\cite{entin:2010} and electron-electron interactions~\cite{hotspots} as possible mechanisms for the transfer of heat (later realized experimentally~\cite{holger,roche:2015,hartmann:2015,dorsch:2020}), followed by configurations exploiting inelastic scattering~\cite{jordan:2013} (realized with quantum dots~\cite{jaliel:2019}), magnons~\cite{sothmann:2012epl} or photons~\cite{ruokola:2012,bergenfeldt:2014,henriet_electrical_2015,antola_quantum_2025}.
The variety of transport configurations considered is large, including capacitively coupled quantum dots~\cite{hotspots,sothmann:2012,sanchez:2013,donsa:2014,whitney:2016,zhang_thermoelectric_2016,dare:2017,walldorf:2017,mayrhofer_stochastic_2021,safdari_effect_2023,ghosh_inverse_2024}, molecules~\cite{entin:2010,entin_three_2012,hwang_proposal_2013,koch_thermoelectric_2014}, nanowires~\cite{bosisio_nanowire_2016}, resonant tunneling~\cite{jordan:2013,sothmann:2013,jiang_enhancing_2014,choi:2015,szukiewicz_optimisation_2016,chiegg_implementation_2017,jiang_nearfield_2018,liu_three_2023}, edge states~\cite{sothmann_quantum_2014,sanchez:2015qhe,chiraldiode,hofer:2015,mani_helical_2018,rouraBas_helical_2018,ndemon,fatemeh,braggio_nonlocal_2024,mishra_reaching_2024}, beam splitters~\cite{genevieve,extrinsic,balduque_coherent_2024,adrian,cioni_high_2025}, quantum dot arrays~\cite{krause:2011,jiang:2012,jiang:2013prb,mazza:2014,arrachea_vibrational_2014,mazza:2015,michalek_local_2016,jiang:2017,lu_unconventional_2021,lu_multitask_2023,cao_quantum_2023}, and tunnel~\cite{ruokola:2012,sanchez_transport_2018,zhou_three_2020}, p-n~\cite{jiang:2013} or Josephson~\cite{hofer:2016} junctions. Three terminal geometries have also been considered for exploring the thermoelectric efficiency bounds when one of the terminals is a probe~\cite{benenti_thermodynamic_2011,sanchez:2011,brandner:2013,balachandran_efficiency_2013,brandner:2013b,brandner_bound_2015,yamamoto_efficiency_2016,sartipi_optimal_2023}. In the case that the resource consists on several terminals, nonthermal configurations~\cite{whitney:2016,ndemon} or thermal drag effects~\cite{lu_electron_2016,bhandari:2018,idrisov_thermal_2022,wang_cycle_2022} can be exploited.
 
We are aware of a few works reviewing related subjects~\cite{bjornreview,benenti:2017,whitney_quantum_2019,wang_inelastic_2022}, however we want to focus on more recent results including phase-coherent effects and experimental realizations, while keeping a global description of the involved physics.
The structure of this review is the following. In Sec.~\ref{sec:multiterm} we present a description of the particle and heat currents and related quantities for multiterminal conductors in general terms. We classify the different configurations in two sections: Sec. \ref{sec:noninteracting} discusses systems using a mean field description of interactions, including inelastic scattering and phase-coherence induced thermoelectric engines, while Sec.~\ref{sec:interact} discusses configurations where the coupling to the heat source is mediated by microscopic interactions. The propagation of heat is the subject of Sec.~\ref{sec:cooling} and perspectives are given in Sec.~\ref{sec:conclusion}.

\section{Multiterminal particle and heat transport}\label{sec:multiterm}

We are interested in configurations as the one depicted in Fig.~\ref{fig:macrothcpl}(c), with multiple (more than two) terminals allowing for particle ($I_i$) and heat ($J_i$) currents, which will be positive when flowing out of their reservoirs, labeled by $i$ or $j$. 
The system is taken out of equilibrium by differences of the reservoir electrochemical potentials, $\mu_i$, and temperatures, $T_i$.
Allowing the system to be composed of electrically isolated regions (i.e., in the case of having capacitively-coupled conductors), labeled by $\alpha$, particle and energy conservation are expressed as $\sum_{i\in\alpha} I_{i}=0$ and $\sum_{i}J_i+\sum_\alpha P_\alpha=0$, where $P_\alpha=-\sum_{i\in\alpha}\mu_iI_{i}$ is the dissipated power in conductor $\alpha$. The system will perform as a thermodynamic engine when power is generated ($P_\alpha>0$) in some conductor $\alpha$, as a refrigerator when heat flows out of a cold reservoir ($J_i>0$), or as a heat pump when heat flows into a hot reservoir ($J_i<0$). In a multiterminal setup, whether a terminal is considered ``hot" or ``cold" depends on its temperature being larger or smaller than a certain reference temperature $T_0$~\cite{manzano_hybrid_2020}. Differently from macroscopic thermocouples though, fluctuating systems that are smaller than the electronic thermalization length may contain regions without a well defined temperature~\cite{whitney:2016}. 

Our discussion is not restricted to the linear response regime, however considering this regime gives useful intuition and will be invoked here when appropriate. Linearizing the particle and heat currents with respect to variations of the electrochemical potentials and temperatures, we get the elements of the Onsager matrix:
\begin{eqnarray}\label{eq:LOWonsagerqd}
\displaystyle
{
\left(\begin{array}{c}
I_i\\
J_i\\
\end{array}  \right)=
\sum_{j}
\left(\begin{array}{cc}
G_{ij} & L_{ij}\\
M_{ij} & K_{ij}\\
\end{array}  \right)
\left(\begin{array}{c}
\mu_j-\mu\\
T_j-T\\
\end{array}  \right)
}
\end{eqnarray}
where $G_{ij}$ is the electrical conductance, and $L_{ij}$ and $M_{ij}$ are the thermoelectric coefficients, related to the Seebeck effect for conversion of heat into power, and to Peltier cooling, respectively. See Fig.~\ref{fig:macrothcpl}(b) for the case of a resonant barrier. The thermal conductance is experimentally defined in the absence of particle currents, hence 
$\mathbf{\tilde{K}}=\mathbf{K}-\mathbf{M}\mathbf{G}^{-1}\mathbf{L}$, where $\mathbf{X}$ are the matrices formed by all $X_{ij}$. From Onsager reciprocity relations, we know that the off-diagonal elements of the full response matrix (the thermoelectric ones, precisely) are related upon the reversal of the magnetic field, $B$: $L_{ij}(B)=M_{ji}(-B)/T$~\cite{onsager1931I,onsager1931II,Casimir1945,jacquod_onsager_2012}. With these, one is able to find the conditions for the different thermoelectric processes to occur under the different boundary conditions.

Multiterminal heat engines have obviously multiple possible resources and multiple possible thermodynamically useful outputs, that can even occur simultaneously~\cite{jiang_enhancing_2014,entin:2015,lu:2017,ndemon,lu:2019,manzano_hybrid_2020,fatemeh,tabatabaei_nonlocal_2022,lu_multitask_2023,lopez_optimal_2023,lu_multitask_2023,tesser_thermodynamic_2023}. Hence, the usual definition of efficiency in terms of the ratio of output power and absorbed heat current needs to be revisited. 
Think for instance in the configuration of Fig.~\ref{fig:macrothcpl}(c): uncareful definitions of the efficiency may diverge when power is generated by a current between terminals $L$ and $R$, while the heat absorbed by the system, $J_H+J_C$, vanishes~\cite{ndemon}.
To have a thermodynamically consistent definition of the engine performance, it is useful to consider the change in the nonequilibrium free energy of the reservoirs, $\dot{F}_i=-I_i\mu_i+J_i(T_0-T_i)/T_i$~\cite{manzano_hybrid_2020,fatemeh}, with the second law imposing that $\sum_i\dot{F}_i\leq0$, i.e., free energy is consumed in dissipation. The efficiency is then written by selecting the positive electric and thermal contributions and dividing them by the negative ones~\cite{manzano_hybrid_2020}:
\begin{equation}
\eta=-\frac{\sum_\alpha^+P_\alpha+\sum_i^+J_i(T_0/T_i-1)}{\sum_\alpha^-P_\alpha+\sum_i^-J_i(T_0/T_i-1)}\leq1,
\end{equation}
where $\sum_i^\pm x_i=\sum(x_i\pm|x_i|)/2$. Generated power is positive, dissipated power is consumed therefore negative. Extracting/injecting heat from/into cold/hot reservoirs (i.e., cooling/pumping) are positive, damped heat is negative.\footnote{Whether a reservoir is considered hot or cold is is established by the reference temperature, $T_0$, which hence has an operational meaning.}
As an example, consider the basic three-terminal thermocouple, with one hot reservoir injecting heat but no particles ($I_H=0$, $J_H>0$) and the other two terminals being at the same temperature ($T_L=T_R=T_0$) and supporting a particle current ($I_L+I_R=0$). Then if power is generated by the current flowing against the electrochemical potential difference, $I_L(\mu_R-\mu_L)$, we recover $\eta=P/(J_H\eta_C)$, with the Carnot efficiency $\eta_C=1-T_0/T_H$. Other definitions of the efficiency are possible based on the rate of entropy change of the different reservoirs~\cite{lu_multitask_2023} that only involve temperatures of the baths.

As transport through nanostructures is highly affected by fluctuations, in the last years it has been emphasized that, on top of the amount of generated power and the efficiency, it is useful to optimize the performance in terms of the constancy of the output flows. In order to reduce the signal to noise ratio, it has been shown that conductors with piecewise constant spectral properties are beneficial~\cite{kheradsoud:2019,timpanaro_hyperaccurate_2023,balduque_coherent_2024} which furthermore improve the efficiency at finite power~\cite{whitney_most_2014}.

\section{Noninteracting electrons}\label{sec:noninteracting}

A simple and elegant description of multiterminal quantum transport is given by scattering theory~\cite{buttiker_scattering_1992}, as long as interactions admit a mean field approach~\cite{landauer_conductance_1989}, see Refs.~\cite{Datta1995,lesovik_scattering_2011,benenti:2017,heikkila_book,moskalets-book} for more details. Reservoirs at different electrochemical potentials $\mu_i$ and temperature $T_i$ inject electrons into the terminals of the conductor, which is treated as a wave scatterer. The scattering matrix ${\cal S}_{ji}(E)$ represents the amplitude for electrons injected with energy $E$ from  terminal $i$ to be transmitted and absorbed by reservoir $j$. 
The initial success of the theory originated in the description of quantized conductance~\cite{vanWees_quantized_1988,wharam_one_1988} and the understanding of the properties and symmetries of the linear response regime~\cite{buttiker_four_1986,benoit_asymmetry_1986,butcher:1990}. However, particle ($I_i$) and heat ($J_i$) currents can be more generally defined in terms of the occupation of the different terminals, given by the fermionic reservoir distribution function, $f_i(E)=1/\{1+\exp[(E-\mu_i)/\kBT_i]\}$:
\begin{align}
\label{eq:Ijscatt}
I_i&=2h^{-1}\sum_{j}
\int{dE}{\cal T}_{ij}(E)[f_i(E)-f_j(E)]\\
\label{eq:Jjscatt}
J_i&=2h^{-1}\sum_{j}
\int{dE}(E-\mu_i){\cal T}_{ij}(E)[f_i(E)-f_j(E)],
\end{align}
where $h$ is the Planck constant, the factor 2 is due to spin degeneracy and ${\cal T}_{ij}(E)=|{\cal S}_{ij}(E)|^2$ is the transmission probability for an electron injected from terminal $j$ and absorbed by terminal $i$.
The unitarity of the scattering matrix preserves particle and energy conservation~\cite{benenti:2017} as expressed in Sec.~\ref{sec:multiterm}. 
With this convention (particle and heat currents are positive when flowing out of reservoirs) and linearizing Eqs.~\eqref{eq:Ijscatt} and \eqref{eq:Jjscatt} when $(\mu_i-\mu)/\kBT$ and $(T_i-T)/T\ll1$, the linear coefficients in Eq.~\eqref{eq:LOWonsagerqd} can be written as $G_{ij}=g_{ij}^{(0)}$, $L_{ij}=g_{ij}^{(1)}/T$, $M_{ij}=g_{ij}^{(1)}$, $K_{ij}=g_{ij}^{(2)}/T$, with~\cite{sivan:1986}
\begin{align}
\label{eq:gijn}
g_{ij}^n=\frac{2}{h}\int{dE}(E-\mu)^n[N_i\delta_{ij}-{\cal T}_{ij}(E)]\left(-\frac{\partial f_{0}(E)}{\partial E}\right),
\end{align}
where $N_i$ is the number of channels of the lead connecting the system to terminal $i$ and $f_{0}(E)$ is the Fermi distribution at equilibrium.
In the limit $\mu\gg\kBT$, these expressions show that the thermoelectric response requires that the scattering problem includes some explicit energy dependence i.e., that the nanostructure acts as a filter to break electron-hole symmetry. If the transmissions are energy-independent, ${\cal T}_{ij}(E)=\tau_{ij}$, one gets $G_{ij}=2(N_i\delta_{ij}-\tau_{ij})/h$ and $K_{ij}=2q_H(N_i\delta_{ij}-\tau_{ij})$, with the quantum of thermal conductance, $q_H=\pi^2\kB^2T/3h$~\cite{pendry_quantum_1983}, for the conductances, and no thermoelectric response: $L_{ij}=M_{ij}=0$. 
The fulfillment of the Onsager reciprocity relations is hence related to the symmetries of the scattering matrix under time reversal satisfying microreversibility: ${\cal S}_{ij}(E,B)={\cal S}_{ji}(E,-B)$~\cite{buttiker_symmetry_1988}, and therefore also ${\cal T}_{ij}(B)={\cal T}_{ji}(-B)$~\cite{butcher:1990}. The linear coefficients plotted in Fig.~\ref{fig:macrothcpl}(b) correspond to a Lorentzian transmission probability usually found in quantum dots~\cite{staring_coulomb_1993,dzurak_observation_1993,svensson_lineshape_2012}. Nonlinear contributions require a careful treatment of the local potentials~\cite{christen_gauge_1996,buttiker_admittance_1997}, determined by particle and entropic injectivities and the onset of interaction effects~\cite{sanchez_scattering_2013,whitney_nonlinear_2013,meair_scattering_2013,whitney_thermodynamic_2013,lopez_nonlinear_2013,texier_nonlinear_2018,sanchez_nonlinear_2019,bipolardiode}.

Let us illustrate the thermoelectric effect with the most relevant case for the remaining discussion: a three-terminal thermocouple, with terminals $L$ and $R$ forming the conductor where current is to be generated, and terminal $H$ on average injecting only heat. The later is described by imposing $H$ to be a voltage probe~\cite{buttiker:1986,buttiker:1988}: i.e., the chemical potential $\mu_H$ will adapt to satisfy the boundary condition $I_H=0$, while having its temperature fixed, $T_H$. 
We can then separate the current in two terms, $I_i=I_{i,el}+I_{i,inel}$, depending on whether electrons flow elastically between the conductor terminals $I_{i,el}\propto\int{dE}{\cal T}_{LR}(E)[f_{L}(E)-f_R(E)]$, or experience inelastic scattering at $H$: $I_{i,inel}=(2/h)\int{dE}{\cal T}_{iH}(E)[f_i(E)-f_H(E)]$. 
Assume for simplicity that the conductor is in equilibrium ($\mu_L=\mu_R$ and $T_L=T_R$), so we write $f_L(E)=f_R(E)=f_0(E)$. Then, transport is fully characterized by the difference of occupation of the heat source and the conductor, $f_H(E)-f_0(E)$, and the scattering probabilities ${\cal T}_{ij}(E)$.
In that case, using $\sum_{i}{\cal T}_{ij}=\sum_{j}{\cal T}_{ij}=1$ and particle conservation, $I\equiv I_L=-I_R$, we are left with only two equations:
\begin{gather}
\begin{aligned}
\label{eq:zbcurrent}
I&=\frac{1}{h}\int dE [\mathcal{T}_{LH}(E)-\mathcal{T}_{RH}(E)][f_0(E)-f_H(E)]\\
0&=\int dE [\mathcal{T}_{HL}(E)+\mathcal{T}_{HR}(E)][f_H(E)-f_0(E)],
\end{aligned}
\end{gather}
corresponding to the generated current and the probe condition, respectively. Note that the elastic transmission, ${\cal T}_{LR}$, does not contribute, and the generated zero bias current is entirely due to the inelastic scattering of electrons at the probe terminal.   
By inspection of Eq.~\eqref{eq:zbcurrent}, we see that for the system to work as a thermocouple able to generate a finite current, $I\neq 0$, it is required that the two transmissions are energy-dependent (in order to generate a thermoelectric response via the broken electron-hole symmetry), and that this dependence also breaks inversion (left-right) symmetry, i.e., $\mathcal{T}_{LH}(E)\neq c\mathcal{T}_{RH}(E)$, with $c$ a positive number. Otherwise, the two integrals become proportional, so the probe condition imposes $I=0$. This implies that the symmetry breaking must be a property of the conductor, not of the coupling to the heat source.  

To generate a finite power, the particle current must flow against a finite bias, $\mu_L\neq\mu_R$. In this case, the elastic transmissions ${\cal T}_{LR}(E)$ and ${\cal T}_{RL}(E)$ introduce contributions that are independent of the heat source (assuming that we can neglect nonlinear effects on the internal potential of the conductor~\cite{buttiker_capacitance_1993}) and hence are detrimental to the device performance~\cite{wang_inelastic_2022}. Therefore power will be enhanced when the heat source is connected in series between the two conductor terminals~\cite{jordan:2013}, so all transported electrons are forced to thermalize at the probe. In analogy with the macroscopic thermocouple, the effect is enhanced if the two contributions to the current in Eq.~\eqref{eq:zbcurrent} have opposite signs, i.e., one is dominated by electrons and the other by holes. This can be done by placing scatterers that separate the heat source from the two conductor terminals having sufficiently sharp spectral features antisymmetric with respect to the electrochemical potential, e.g., quantum dots~\cite{jordan:2013}. It was found that in order to optimize the efficiency for a given power, boxcar shaped transmissions are ideal~\cite{whitney_quantum_2016}, also for nonthermal resources~\cite{danielsson_optimizing_2025}. Finding systems with these properties is unfortunately not easy, so alternatives are requested~\cite{chrirou_potential_2025}. We discuss below the different strategies proposed so far.
To ease the discussion, we will restrict to single-channel and time reversal symmetric conductors in the following, except when explicitly including a magnetic field.

\subsection{Inelastic junctions}\label{subsec:inelastic}

\begin{figure*}[t]
\centering
\includegraphics[width=\textwidth]{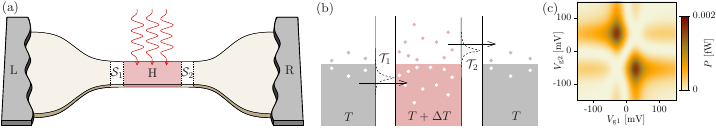}
\caption{(a) Inelastic-scattering junction, with the region in contact with the heat source sandwiched between two scattering regions ${\cal S}_{\alpha}$, $\alpha=1,2$, which connect it with the conductor terminals $L$ and $R$. (b) Resonant tunneling barriers introduce Lorentzian-shaped filters ${\cal T}_{\alpha}(E)$ that filter the transported electrons. Resonances can be tuned with gate voltages $V_{g\alpha}$. Those below (over) the electrochemical potential induce transport from hot to cold (from cold to hot) regions. (c) A finite power $P$ is generated in asymmetric configurations by generating a current through a load resistor closing the circuit. Here $R_{load}\gg R_K$, with resonances of width $\Gamma_1=\unit[0.2]{meV}$, $\Gamma_1=\unit[0.1]{meV}$, $T=\unit[175]{mK}$, $\Delta T=\unit[5]{mK}$, and being $R_K=h/e^2\approx\unit[25.813]{k\Omega}$ the von Klitzing resistance.
}\label{fig:inelast}
\end{figure*}

Let us start by assuming that the undesired elastic transmission between $L$ and $R$ terminals is neglected, ${\cal T}_{LR}\ll1$. Electrons are then forced to thermalize with the heat source (therefore maximizing the exchange of heat) before being transmitted through the conductor, as in the case depicted in Fig.~\ref{fig:inelast}(a). This can be done by sandwiching the region in contact with the heat source between two scattering regions that can be tuned via gate voltages, or by coupling the mesoscopic region to a hot phonon bath. While some of these configurations have been reviewed before~\cite{bjornreview,wang_inelastic_2022}, we focus here on some more recent contributions. 

A way to model the inelastic region is by introducing a probe terminal, say $H$, where electrons thermalize at a thermal distribution defined by a given temperature $T_H=T+\Delta T$ and an electrochemical potential, $\mu_H$, that is obtained from the boundary condition of a voltage probe~\cite{engquist_definition_1981,buttiker:1988}, $I_H=0$.
For the simple case of a hot inelastic region sandwiched by two scattering regions in a two terminal conductor, see Fig.~\ref{fig:inelast}(a), the linear regime gives a current depending on the response coefficients of the individual junctions:
\begin{equation}
\label{eq:inelasticlinear}
I_L=\frac{G_{RH}L_{LH}-G_{LH}L_{RH}}{G_{LH}+G_{RH}}\Delta T,
\end{equation}
with a developed potential $\mu_H-\mu=-(L_{LH}+L_{RH})\Delta T_H/(G_{LH}+G_{RH})$.\footnote{The two junctions act as two parallel resistors to the particle current flowing from the hot terminal generated by them.} As shown in Fig.~\ref{fig:macrothcpl}(b), $L_{LH}$ and $L_{RH}$ of opposite signs can be achieved for resonant barrier scatterers by tuning gate voltages such that they can both contribute positively to the current in Eq.~\eqref{eq:inelasticlinear}.

\subsubsection{Resonant tunneling energy harvesters}\label{sec:resonant}

To overcome the low power generated in previously investigated Coulomb-blockade heat engines (see Sec.~\ref{sec:interact}), resonant tunneling barriers where proposed~\cite{jordan:2013} which allow for strong coupling to the heat source at the same time enabling the energy filtering to have an efficient thermoelectric response~\cite{humphrey:02}. A single channel conductor was proposed consisting on quantum dots connecting the conductor terminals with a cavity where electrons thermalize at an increased temperature, see Fig.~\ref{fig:inelast}(b). Therefore, the setup can be modeled as electrons tunneling with probability~\cite{buttiker:1988}
\begin{equation}
\label{eq:lorentz}
{\cal T}_{Hj}(E)=\frac{\Gamma_{l_ja}\Gamma_{l_jb}}{(E-\varepsilon_{l_j}^{})^2+(\Gamma_{l_ja}+\Gamma_{l_jb})^2/4}
\end{equation}
between three reservoirs in series, the outer ones ($j=L,R$) at a given temperature $T$ and electrochemical potentials $\mu_L$ and $\mu_R$, the center one, $H$, being hotter $T_H=T+\Delta T$, and with $\mu_H$ resulting from particle conservation, $I_H=0$, and therefore given by the sum of (two-terminal) thermovoltages of the two separated junctions. In Eq.~\eqref{eq:lorentz}, $\varepsilon_{l_j}^{}$ is the energy of the quantum dot connecting $H$ with $j$ ($l_L=1$, $l_R=2$), and $\Gamma_{l_jk}/\hbar$ the tunneling rates of the two barriers forming them.

The quantum dot energies, $\varepsilon_l$, can be tuned with gate voltages, one of them to filter particles below the electrochemical potential (hence contributing to transport into the hot cavity), the other one over it (therefore extracting particles from the cavity)~\cite{jordan:2013,szukiewicz_optimisation_2016}, which results in a direct current that can power a load resistor, as shown in Fig.~\ref{fig:inelast}(c). 
The generated current can be scaled up by parallelization in a layered structured of self-assembled quantum dots~\cite{jordan:2013}. Alternatively, quantum wells~\cite{sothmann:2013} or superlattices~\cite{choi:2015} enhance the available currents while keeping the filtering properties (the maximum power is $P\sim\unit[0.1-0.3]{W/cm^2}$ at an efficiency around $0.1-0.2\eta_C$ at room temperature~\cite{choi:2015}). 
The band quantization of one-dimensional wires~\cite{yang_optimal_2020} and two-dimensional heterostrusctures~\cite{yang_three_2020} can be used as low-pass filters as mesoscopic versions of thermionic devices, which however compromise the efficiency. 
Multiple resonance barriers~\cite{liu_three_2023} have also been considered which may approach the optimal configuration, with boxcar-shaped transmissions, in terms of efficiency at finite power~\cite{whitney_quantum_2016}. The three-terminal composition can also be exploited to enhance the engine performance by allowing for cooperative effects due to having two temperature differences ($T_L\neq T_R\neq T_H$)~\cite{jiang_enhancing_2014}.

A proof of concept experiment was realized by defining an electronic cavity coupled to two electronic terminals via single quantum dots in a semiconductor 2DEG~\cite{jaliel:2019}. The cavity is additionally connected to a heating channel allowing its temperature to increase by around $\unit[55]{mK}$ with respect to the conductor terminal temperature $T\sim\unit[85]{mK}$. The asymmetry of the device is controlled by tunning gate electrodes connected to the two dots, resulting in a generated current loading a $\unit[0-2]{M\Omega}$ resistor, as modeled in Fig.~\ref{fig:inelast}(c). A maximum power $P\approx\unit[0.12]{fW}$ is measured for $R_{load}\approx\unit[1]{M\Omega}$. The estimated efficiency is however smaller than the ideally predicted $\eta_C/2$. This is however not a fundamental limitation, as quantum dots properly isolated from their environment (e.g., by defining them in suspended wires) have demonstrated a high efficiency of close to the Carnot bound in similar temperature regimes~\cite{josefsson_quantum_2018}.

\subsubsection{Coupling to phonons}\label{sec:phonons}

The natural source of inelastic scattering in a nanostructure is via the coupling to phonons. Phonons and electrons are typically at different temperatures, so the problem is equivalent to a three-terminal configuration, with the phonon bath acting as the heat source. For this coupling to be relevant and able to induce a thermoelectric response via the conversion of heat exchanged in the conductor, one needs to assume that the action of the phonons is localized in the nanostructure. A way to do this is by using near-field absorbers sandwiched in an energy-selective layer structure similar to those discussed in the previous section: the electrons excited by phonon absorption tunnel to the collector through a high-energy filter (e.g., self-assembled quantum dots). The low-energy holes left behind are filled by electrons tunneling from the emitter reservoir~\cite{jiang_enhancing_2014,wang_hot_2023}, bearing similarities with hot-carrier solar cells~\cite{tesser_thermodynamic_2023}. 

Inhomogeneous electron-phonon coupling is also found in hybrid normal-superconductor nanostructures and in tunnel junctions. In the former case, the electron-phonon coupling is negligible in the superconductor. 
A normal island connected to two superconducting reservoirs will hence develop a temperature difference with respect to these. The gap of the superconductors can then be used as filters enabling the thermoelectric response. 
A direct current will be generated by converting the heat exchanged with the phonons, for which the two gaps need to be different~\cite{donald}. In tunnel junctions, one can use the volume dependence of the electron-phonon coupling, such that islands of different sizes will exchange different amounts of heat with the phonon bath. The temperature profile along the conductor can be considered piecewise constant between tunneling barriers, which results in broken detailed balance (and hence in a thermoelectric current) when combined with gating, even if all barriers are energy independent~\cite{sanchez_transport_2018}.     

Another possibility is when the nanostructure contains a barrier or a bottleneck transition that the energy gained by absorbing a phonon helps to overcome~\cite{entin:2010}. This case occurs naturally in lattices of localized states such as molecules~\cite{entin:2010}, quantum dots~\cite{jiang:2012} or suspended wires~\cite{jiang:2013prb,bosisio_using_2015,bosisio_nanowire_2016}. We discuss the interaction of electrons with single phonons in Sec.~\ref{sec:ephon}.

\subsection{Phase coherent thermocouples}\label{subsec:3t}

\begin{figure*}[t]
\centering
\includegraphics[width=\textwidth]{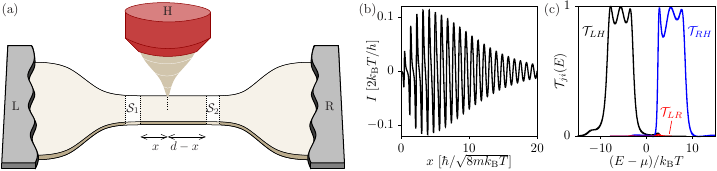}
\caption{Phase coherent coupling to the heat source. (a) A scanning probe tip injects heat at a given position $x$ between two scattering regions ${\cal S}_{\alpha}$, $\alpha=1,2$, separated by a distance $d$. (b) Extrinsic thermoelectric response of a conductor consisting on a single energy-independent scatterer, ${\cal S}_1$, with transmission probability ${\cal T}_1=0.3$ and ${\cal S}_2=\mathbb{1}$: the interference of multiple reflections between the scatterer and the tip generates a finite current (here, $T_H=1.5T$, the conduction band bottom is $U_0=-50\kBT$, and the tip is maximally coupled, $\epsilon=1/2$, see Ref.~\cite{extrinsic} for details). (c) Transmission probabilities between the different terminals when the two scatterers are double quantum dots with transparencies $3.1\kBT$, interdot coupling $2.1\kBT$ and splitting $11.7\kBT$, with the tip with transmission $\epsilon=0.4$ at $x=6.2l_0$, being $d=6.4l_0$ and $l_0=\hbar/\sqrt{8m\kBT}$, with the electron mass $m$~\cite{balduque_coherent_2024}.
}\label{fig:elast}
\end{figure*}

Differently from the inelastic configurations discussed in Sec.~\ref{subsec:inelastic}, in fully coherent three-terminal junctions, the detrimental contribution of the longitudinal transmission probability ${\cal T}_{LR}(E)$ cannot be neglected in general. 
However, this allows us to handily introduce the phase of the electron waves as a new degree of freedom that can be harnessed to provide the system with the features required to work as a thermocouple, tune its thermoelectric properties, or boost its performance.

To show this, it is useful to partition the junction in three regions: two scattering  elements, $\mathcal{S}_1$ and $\mathcal{S}_2$, mediating the transport between the junction and the isothermal terminals $L$ and $R$, and a central region where the conductor couples to the heat source, which is injected coherently into the scattering region. A possible realization is sketched in Fig. \ref{fig:elast}(a), where the heat source is connected to a scanning probe tip allowing for the local injection of heat at a chosen position and for choosing the coupling strength (and even be disconnected from the conductor). 
The electron waves scattered into the channels connecting two consecutive scattering elements accumulate a phase that depends on the distances $x$ and $d$ which, after undergoing multiple reflections between them, gives rise to interference processes. This phase can be furthermore controlled: kinetic phases, which are proportional to $k(E)=\sqrt{2m(E-U_0)}/\hbar$, are tuned with plunger gates acting on the base electrostatic potential, $U_0$; magnetic phases in annular configurations are tuned via the magnetic flux piercing the ring, $\Phi$~\cite{gefen_quantum_1984,buttiker:1984}.

Equation~\eqref{eq:inelasticlinear} shows that inelastic thermocouples need to be formed of elements with a finite Seebeck response. A remarkable difference in phase coherent junctions is the possibility to use the hot terminal to induce an {\it extrinsic} thermoelectric effect in a conductor that would exhibit no thermoelectricity without the probe~\cite{extrinsic}. A simple example is a one channel conductor with an energy-independent scattering matrix ${\cal S}_1$ (a barrier) connecting terminals $L$ and $R$ to which a hot scanning tip is approached. This is the configuration of Fig.~\ref{fig:elast}(a) with ${\cal S}_2=\mathbb{1}$ and ${\cal S}_1$ depending on a constant transmission probability.
The kinetic phase accumulated in multiple reflections between the barrier and the tip (separated by a distance $x$) introduces an energy dependent $2k(E)|x|$-periodic interference pattern in the three-terminal scattering matrix (composed of the barrier and the tip) sufficient to break the electron-hole symmetry~\cite{buttiker:1989}. Meanwhile, the inversion symmetry is broken geometrically by the position of the tip with respect to the barrier.
The interference pattern can be further tuned by modifying the position of the tip. 
This results in a tunable thermoelectric effect as shown in Fig. \ref{fig:elast}(b), where the current, $I=I_L=-I_R$, exhibits a series of oscillations with $x$. A similar non-local oscillating behavior was found in tight-binding simulations of a quantum point contact acting as the isothermal conductor~\cite{genevieve}. The tunability of the transport properties of these systems allow furthermore for highly efficient thermoelectric rectification~\cite{genevieve,extrinsic}. Unfortunately, the heat to work conversion performance is low in these devices, as a consequence of a considerable fraction of the heat being lost to the terminal without thermoelectric scatterer and the transmission probability being sub-optimal.

The performance can be improved by considering scattering regions with more useful properties ${\cal S}_\alpha(E)$ and exploring the capability to engineer the global transmission probabilities of a quantum thermocouple using the electron phase. A system comprised by quantum dots (or double quantum dots) as scattering regions $1$ and $2$, has been proposed~\cite{balduque_coherent_2024} as a coherent analog of the inelastic thermocouples~\cite{jordan:2013,jaliel:2019} discussed in Sec.~\ref{sec:resonant}. While this system allows for direct transport between terminals $L$ and $R$, which is in principle detrimental for the thermocouple operation (as discussed in Sec. \ref{sec:noninteracting}), optimizing the different available parameters one finds regimes for which the tip-mediated interference within the conductor becomes destructive and ${\cal T}_{LR}(E)\approx0$, see Fig. \ref{fig:elast}(c). Furthermore, this interference can improve the transmission properties of the two junctions giving rise to an enhancement of all performance quantifiers with respect to the inelastic case. In particular, for double quantum dots, the resulting transmissions develop a boxcar-shape~\cite{whitney_quantum_2016}, attaining efficiencies of $\eta > 0.4\eta_C$ with high extracted powers (around $70\%$ of the theoretical maximum) and reduced signal to noise ratio~\cite{balduque_coherent_2024}. A slight improvement of the performance was also found in tight-binding coherent models of the quantum dot thermocouple configuration just discussed~\cite{chiegg_implementation_2017}, which can again be attributed to suppressed transport between isothermal terminals and the favorable modification of the $H-L$ and $H-R$ transmissions.

The three terminal thermoelectric response in phase coherent setups has also been proposed for the detection of helical states in topological Josephson junctions~\cite{blasi_nonlocal_2020,blasi_nonlocal_2021,mateos_nonlocal_2024}: the coupling to a normal probe affects the dispersion of otherwise electron-hole symmetric edge states connecting two superconductors, inducing a finite longitudinal current between them. 

The manipulation of the electron phase via the Aharonov-Bohm effect~\cite{aharonovbohm} in three-terminal coherent ring configurations to design quantum thermocouples has also been explored~\cite{entin_three_2012,hwang_proposal_2013,Behera2023Oct,adrian}. The transport properties of these devices are easily tunable via the magnetic phase $\phi=2\pi\Phi/\Phi_0$ accumulated upon performing a loop around the ring, with $\Phi$ being the magnetic flux piercing the ring and $\Phi_0=h/e$ the flux quantum~\cite{washburn_aharonov_1986,ihn_semiconductor_2009}. 
In these setups the electron-hole symmetry is broken by the kinetic phase accumulated between the junctions connecting the ring to the different terminals~\cite{guttman_thermopower_1995,haack:2019}, with broken time-reversal symmetry introducing the mirror asymmetry (via the non reciprocity of the transmission probabilities ${\cal T}_{ij}(B)\neq{\cal T}_{ji}(B)$) needed for the thermocouple effect, even if the geometric configuration of the system is symmetric and energy-independent~\cite{adrian}. To see this, consider a symmetric ring with all three terminals equally separated by one third of the ring perimeter, so we have ${\cal T}_{\circlearrowright}(\phi)={\cal T}_{12}(\Phi)={\cal T}_{23}(\Phi)={\cal T}_{31}(\Phi)$. 
A Sommerfeld expansion~\cite{sommerfeld:1928} of the linear response coefficients (valid at low temperatures, $\kBT\ll\mu$ provided the transmission probabilities are smooth around $E=\mu$~\cite{extrinsic}), such that $I_i\approx2\sum_{j}[(h^{-1}{\cal T}_{ij}(\mu)(\mu_i-\mu_j)+q_H{\cal T}'_{ij}(\mu)(T_i-T_j)]$, results in a zero-bias current:
\begin{equation}
I_1=2q_H\frac{{\cal T}'_{\circlearrowright}(\mu,\phi){\cal T}_{\circlearrowright}(\mu,{-}\phi){-}{\cal T}'_{\circlearrowright}(\mu,{-}\phi){\cal T}_{\circlearrowright}(\mu,\phi)}{{\cal T}_{\circlearrowright}(\mu,\phi)+{\cal T}_{\circlearrowright}(\mu,-\phi)}\Delta T_H,
\end{equation}
where ${\cal T}'_{\circlearrowright}=\partial_E{\cal T}_{\circlearrowright}$.
However, the direct elastic channel between the isothermal terminals poses a drawback for their optimization as thermocouples.

\subsubsection{Dephasing}
As all of these proposals rely in the phase-coherence of the electrons within the conductor, the presence of dephasing or thermalization mechanisms imposes a limitation in their performance. The impact of different forms of dephasing has been studied by introducing virtual probes to the problem, where electrons propagating along the conductor can be absorbed and reinjected, resulting in phase randomization~\cite{buttiker:1986,buttiker:1988,dejong_semiclassical_1996}. The particular characteristics of each probe allow to model different scenarios such as pure dephasing without momentum backscattering, quasielastic scattering or the thermalization of the electrons \cite{extrinsic,genevieve,adrian}. When quantum interference is the only responsible of the electron-hole energy breaking, the nonlocal thermoelectric effect is indeed affected by dephasing and disappears when all electrons lose their phase in their way through the conductor. 

The relaxation of electrons within the conductor can also be modeled via probe terminals by imposing the boundary conditions $I_p=J_p=0$: the probe adapts its temperature and electrochemical potential, thus behaving as a thermometer, and enables the definition of an effective local temperature in different regions of the conductor~\cite{jacquet_temperature_2012,bergfield_probing_2013,meair:2014,shastry_temperature_2016,zhang_local_2019,shastry_scanning_2020}, which also shows an interference pattern~\cite{extrinsic}. The measurement of local dissipation in one dimensional conductors via scanning probe tips has also been described with more involved models combining the scattering approach with a nonlinear Boltzmann equation for the electrons \cite{leumer_going_2024}.

\subsection{Edge states}\label{subsec:chiral}

\begin{figure*}[t]
\centering
\includegraphics[width=\textwidth]{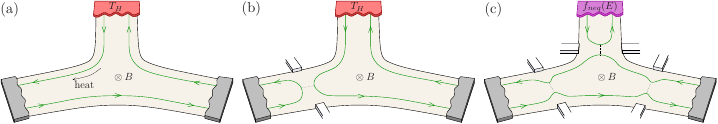}
\caption{Three terminal quantum Hall conductor. (a) In the absence of backscattering, heat flows from the hot probe terminal into the next terminal along the chiral propagation of electrons, with no average currents $I_i=0$. (b) A scatterer mixing edge channels is able to generate a thermoelectric response. (c) A nonthermal reservoir injects electrons in one of the channels with a nonequilibrium distribution $f_{neq}(E)$ able to generate a finite current even if $I_{neq}=J_{neq}=0$.   
}\label{fig:chiral}
\end{figure*}

A way to control which electrons get in contact with the heat source is to use the chirality of quantum Hall edge states~\cite{klitzing_new_1980,buttiker_absence_1988} in the presence of strong magnetic fields. The quantum Hall regime is indeed a multiterminal effect, and thermoelectric measurements have elegantly demonstrated the chiral propagation of electrons via the energy dissipated in the terminals along the way from the heat source~\cite{granger_observation_2009,nam_thermoelectric_2013,huynh_chiral_2025}. Measurements of the temperature of a small (a few $\unit[]{\mu m^2}$) metallic island coupled via a selected number of edge channels to a dissipative current evidenced the quantization of the thermal conductance~\cite{kane_quantized_1997,jezouin_quantum_2013} and the occurrence of heat Coulomb blockade if the island $RC$ time does not allow to stabilize the charge fluctuations~\cite{slobodeniuk_equilibration_2013,sivre_heat_2018}.

In addition to the separation of heat and particle flows discussed in the previous sections, the quantum Hall regime achieves the separation of left moving and right moving electrons on opposite sides of the conductor, making it possible to inject heat only on electrons moving in a particular direction~\cite{aita_heat_2013,sothmann_quantum_2014,sanchez:2015qhe,chiraldiode,sanchez_effect_2016}, see Fig.~\ref{fig:chiral}(a). 
In the absence of interedge scattering, heat flows only along the edge connected to the hot terminal. Along the other edge, the particle current vanishes. In order to generate a finite thermoelectric response, the two edges need to be coupled by an energy-dependent scattering region, see Fig.~\ref{fig:chiral}(b). This can be a gate tunable junction like, e.g., a quantum point contact or a quantum dot, or induced by the difference of kinetic phases in a Mach-Zehnder interferometer~\cite{hofer:2015}. Interestingly, if the scatterer is placed only between the heat source and one of the conductor terminals, as shown in Fig.~\ref{fig:chiral}(b), the Seebeck and Peltier effects (which are linked by Onsager reciprocity relations in the absence of a magnetic field) are also separated, i.e.,  one is finite with the other one vanishing~\cite{sanchez:2015qhe}: $\Delta T$ generates a current between $L$ and $R$ (hot particles from $H$ are filtered at the scatterer), but a difference in $\mu_L-\mu_R$ does not generate a thermoelectric heat current in terminal $H$ (particles injected from $L$ never reach $H$). This effect was related to a possible enhancement of the efficiency at maximum power~\cite{benenti_thermodynamic_2011}, however it was later shown that the balance established by the not scattered channel restores the usual bounds~\cite{whitney_quantum_2016,mishra_reaching_2024}.

The same discussion can be applied to helical states in topological insulators, which preserve time-reversal symmetry but separate electrons with opposite spin polarization. In that case, the injected heat is converted into a pure spin current~\cite{sanchez:2015qhe,mani_helical_2018,rouraBas_helical_2018}. 

The hot terminal can be replaced by a driven resonator achieving the ultrastrong coupling of terahertz radiation and electrons in an edge state. The injected polaritons (relaxing as electron-hole excitations) give rise to a chiral photocurrent~\cite{huang_chiral_2025}. 

\subsection{Nonequilibrium states}\label{subsec:neq}

In the configurations discussed above, the thermocouple was able to generate a finite electrical power out of an heat current from a hotter (or colder) thermal bath.  
One may think also on configurations where the thermodynamic resource is not a thermal reservoir in local equilibrium, i.e., not described by a well defined temperature and electrochemical potential but rather by a nonequilibrium distribution. In that case, the thermocouple can take advantage of the nonequilibrium state provided its size is smaller than the electron thermalization length~\cite{whitney:2016}. In this case, the usual limitations imposed by the thermodynamic laws (holding for flows between equilibrium baths) can be chased. For instance, configurations can be found where finite power $P>0$ is generated in the conductor terminals without any net particle or heat flow from the resource terminal(s)~\cite{ndemon}. The conductor currents then fulfill the conservation laws establishing the first law of thermodynamics,  $I_L+I_R=0$ and $J_L+J_R=-P$, while its entropy decreases: $\dot S_s=\dot S_L+\dot S_R=-P/T$, hence apparently violating the second law~\cite{ndemon}. The nonequilibrium environment hence has the same effect on the average system properties as a Maxwell demon would have~\cite{whitney_illusory_2023}, autonomously and with no need to extract information of microscopic states. Of course, the global system must satisfy the second law, which can be verified by considering the entropy currents~\cite{deghi_entropy_2020} or the consumed nonequilibrium free energy of the resource~\cite{fatemeh}. The action of the demon on the conductor currents needs to be tracked in their fluctuations~\cite{freitas_characterizing_2021}: it has been shown that useful thermodynamic processes require a minimal amount of noise~\cite{acciai_constraints_2024}.

Proposals to observe such effect include the use of quantum Hall edge states that facilitate the preparation of a nonequilibrium distribution: two terminals at different local states (e.g., $f_C(E)$ and $f_H(E)$) can be connected by a scatterer which mixes their distribution. Electrons in one of the outgoing channels from this scattering region will have a distribution $f_{neq}(E)={\cal T}_C(E)f_C(E)+{\cal T}_H(E)f_H(E)$, where ${\cal T}_i$ is the transmission probability for electrons injected from terminal $i$~\cite{ndemon,fatemeh}. Electrons in the edge state with the resulting distribution can then be injected into the conductor inducing a finite power if they are scattered at a junction with a thermoelectric response, similarly as the discussion in Sec.~\ref{subsec:chiral}, see Fig.~\ref{fig:chiral}(c). 
In this model, the fulfillment of the second law can be related to the dissipative currents within the demon, $J_H=-J_C$, with $\dot{S}_d=\dot{S}_C+\dot{S}_H=J_H(T_C^{-1}-T_H^{-1})\geq-\dot S_s$. The nonequilibrium situation can also be generated by time-dependent pulses~\cite{ryu_beating_2022}. The addition of channels coupled to different phonon baths have also been discussed to have a similar effect that however cannot be related to a single nonequilibrium state~\cite{lu_unconventional_2021}.

In experimental configurations, the nonequilibrium situation can be a result of different processes acting on different time scales. This is precisely the regime of operation of hot-carrier solar cells, which can be modeled as a multiterminal conductor~\cite{tesser_thermodynamic_2023,bertinJohannet_improving_2025}. Nonthermal distributions can also be found in interacting edge channels~\cite{altimiras_nonequilibrium_2010,lesueur_energy_2010,itoh_signatures_2018,fujisawa_nonequilibrium_2022} and have been shown to improve the thermoelectric efficiency~\cite{yamazaki_efficient_2025}.
Mesoscopic conductors are then of help for understanding how to extract useful power out of nonequilibrium states before they relax in practical setups.

\section{Interacting systems}\label{sec:interact}

\begin{figure*}[t]
\centering
\includegraphics[width=\textwidth]{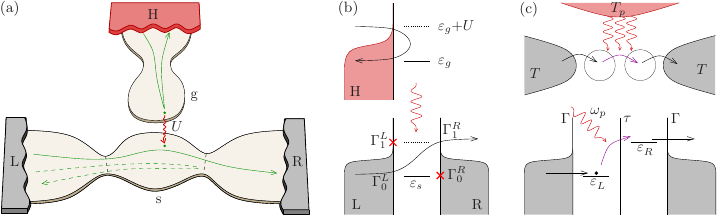}
\caption{Nonlocal thermoelectrics mediated by interactions. (a) The heat injection is mediated by the Coulomb interaction, $U$, of electrons in capacitively coupled conductors. The coupling of the nanostructure to the $L$ and $R$ reservoirs needs to be energy-dependent and inversion asymmetric. (b) Sequence leading to a finite zero-voltage current with quantum dots with asymmetric tunneling rates $\Gamma_{n}^i$. (c) Phonon assisted tunneling between two quantum dots provides the energy, $\hbar\omega_p\approx\varepsilon_R-\varepsilon_L\gg\lambda$ to overcome an energy gap, where $\lambda$ is the interdot coupling, and $\Gamma$ are the tunneling rates for coupling with the reservoirs. The splitting of the quantum dot levels introduce all the required broken symmetries naturally.
}\label{fig:interact}
\end{figure*}

In some quantum systems, single-particle interactions become important and cannot be treated in a mean-field level. This is the case of quantum dot systems weakly coupled to electronic reservoirs~\cite{vanderWiel_electron_2002}, where strong Coulomb interactions stabilize the number of allowed electrons (Coulomb blockade regime~\cite{grabert_single_book}) up to single-charge fluctuations through the tunnel barriers~\cite{beenakker_theory_1992}. 
For example, Coulomb interactions between electrons in capacitively coupled conductors or interactions with bosonic excitations (phonons, photons, magnons) need to be described microscopically. 
Existing proposals discussed in this section use systems designed such that a bottleneck transition for transport is suppressed but can be overcome via one of these interactions, this way correlating the particle transport with the transfer of heat quanta, see Fig.~\ref{fig:interact}. 

When the system-bath couplings are weak, these systems fall in the category of open quantum systems~\cite{breuer:book,schaller:book,philipp_book}. A simple description of their dynamics is given in terms of the master equation for the evolution of the system (reduced) density matrix, $\hat\rho(t)$. One usually assumes Born-Markov and secular approximations, resulting in the Gorini-Kossakowski-Sudarshan-Lindblad master equation~\cite{gorini_completely_1976,lindblad_generators_1976,manzano_short_2020}
\begin{equation}
\label{eq:lindbladeq}
\dot{\hat\rho}=
-\frac{i}{\hbar}[\hat H_S,\hat\rho]+\sum_{is}\sum_{qp}W_{qp}^{is}\left(\hat Y_{i,qp}^{}\hat\rho \hat Y_{i,qp}^{\dagger}-\frac{1}{2}\{\hat Y_{i,qp}^\dagger \hat Y_{i,qp}^{},\hat\rho\}\right)
\end{equation}
where the first term in the right hand side represents the coherent dynamics induced by the system Hamiltonian, $\hat H_S$, and the second term describes the dissipative dynamics induced by the reservoirs,
produced by jump operators $\hat Y_{i,qp}$ describing transitions between the system states $|p\rangle\rightarrow|q\rangle$ due to the system-bath interaction $H_{SB}$. For the coupling to the electronic reservoir $i$, this is $\hat{H}_{SB,i}=\sum_{k_i}\lambda_i\hat{d}_{\alpha_i}^\dagger\hat{c}_{k_i}^{}+{\rm H.c.}$, where $\lambda_i$ is the weak system-bath coupling, $\hat{d}_{\alpha_i}^\dagger$ creates an electron in quantum dot $\alpha_i$ and $\hat{c}_{k_i}^{}$ destroys it in reservoir $i$. The transition rates $W_{qp}^{is}$ are given by the Fermi golden rule, where the index $s=\pm$ labels whether the transition involves a particle tunneling into/out of the system~\cite{schaller:book}: 
\begin{align}
W_{qp}^{i+}&=\frac{2\pi}{\hbar}\left|\lambda_i\langle q|\hat{d}_{\alpha_i}^\dagger|p\rangle\right|^2\nu_i(\Delta_{qp})f_i(\Delta_{qp})\\
W_{qp}^{i-}&=\frac{2\pi}{\hbar}\left|\lambda_i\langle q|\hat{d}_{\alpha_i}^{}|p\rangle\right|^2\nu_i(-\Delta_{qp})[1-f_i(-\Delta_{qp})],
\end{align}
where $\nu_i(E)$ is the density of states in the reservoir and $\Delta_{qp}=E_q-E_p$ is the energy difference of the two system states.
The average particle and heat currents out of the reservoirs are then 
\begin{align}
\label{eq:meI}
I_i&={\rm tr}\Big({\sum_{is}sW_{qp}^{is}\hat Y_{i,qp}\hat\rho}\Big)\\
\label{eq:meJ}
J_i&={\rm tr}\Big({\sum_{is}(E_p-E_q)sW_{qp}^{is}\hat Y_{i,qp}\hat\rho}\Big)-\mu_iI_i.
\end{align}
In the later equation, it is assumed that the energy levels are discrete, which is a good approximation in the weak coupling regime, $W_{qp}^{is}\ll\kBT_i/\hbar$.
The dc currents are obtained by simply replacing the stationary solution of the master equation, $\dot{\hat\rho}=0$, in the above expressions.
Regimes with stronger couplings require more involved treatments, e.g., via higher order perturbative expansions~\cite{konig_resonant_1996,karlstrom_diagrammatic_2013} or in terms of nonequilibrium Green functions~\cite{jauho_book,jaimealfredo}.

\subsection{Electron-electron interactions}\label{subsec:ee}

Capacitively coupled quantum dots (one in the two-terminal conductor system, $s$, and the other one, $g$, coupled to a hot terminal, as in Fig.~\ref{fig:interact}(a)) enable the spacial separation of the heat source and the conductor, with the heat exchange being restricted to the mesoscopic region. Electrons in different dots interact via the Hamiltonian $\hat H_U=U\hat{n}_s\hat{n}_g$, where $U$ is the interdot Coulomb repulsion and $\hat{n}_\alpha=\hat{d}_\alpha^\dagger\hat{d}_{\alpha}^{}$ is the occupation of dot $\alpha$, such that a series of fluctuations 
\begin{equation}
\label{eq:sequence}
(n_s,n_g)=(0,0)\rightarrow(1,0)\rightarrow(1,1)\rightarrow(0,1)\rightarrow(0,0), 
\end{equation}
illustrated in Figs.~\ref{fig:interact}(a) and \ref{fig:interact}(b), transfers an energy quantum $U$ from $g$ to $s$. This energy is provided by the hot reservoir coupled to the former dot. When the two system transitions $(0,0)\rightarrow(1,0)$ and $(1,1)\rightarrow(0,1)$ occur through different barriers coupling dot $s$ to terminals $L$ and $R$, the sequence results in an electron transported through the conductor, enabling the heat conversion engine~\cite{hotspots}. 
This sequence is relevant when the transport of an electron without interacting with $g$, $(0,0)\rightarrow(1,0)\rightarrow(0,0)$, is suppressed, e.g., if $\Gamma_0^R\ll\Gamma_0^L$, with $\Gamma_{n_g}^i$ being the rate for tunneling events through the barrier $i$ between states $(0,n_g)$ and $(1,n_g)$. In this case, the absorption of an energy $U$ activates the transition with a larger rate $\Gamma_1^{R}$, see Fig.~\ref{fig:interact}(b). For the process to generate a finite current, the tunneling barriers need to be left-right asymmetric and energy dependent: in the sequential tunneling regime, the current is
\begin{equation}
\label{eq:currhotspot}
I(\Delta\mu=0)\propto(\Gamma_{0}^L\Gamma_{1}^R-\Gamma_{0}^R\Gamma_{1}^L)J_H/U\sim[f_H(E)-f_0(E)]. 
\end{equation}
This effect has been confirmed experimentally in various settings of coupled quantum dots~\cite{holger} and open cavities~\cite{roche:2015,hartmann:2015}: A small current $\sim$\unit[0.2--3]{pA} is generated which can be reversed by gating the tunneling asymmetry~\cite{holger,roche:2015} at \unit[0.2--0.8]{K} temperatures. 
Similar processes lead to the mesoscopic Coulomb drag effect in voltage driven sources~\cite{Moldoveanu2009Jul,sanchez:2010,kaasbjerg:2016,sierra:2019,bischoff:2015,keller:2016} and thermal drag in temperature biased coupled systems~\cite{bhandari:2018,wang_cycle_2022}. If the two rates in either of the terms contributing to the current are negligible (e.g., by including additional energy filters~\cite{hotspots,strasberg:2018}), all electrons that absorb an  energy $U$ move in the same direction, increasing power and leading to a Carnot efficient heat engine~\cite{hotspots}, see also reviews discussing this configuration for more details~\cite{bjornreview,thierschmann_thermoelectrics_2016,benenti:2017,cangemi_quantum_2024}. 

In the sequential tunneling regime just discussed, with $\hbar\Gamma_n^i\ll\kBT_i$, phase coherence is lost in every transition. Transport is strongly dominated by nonequilibrium fluctuations, which can be traced in time via charge detectors~\cite{fujisawa:2006,kung_irreversibility_2012}. This makes not only the particle but also the transferred heat statistics accessible~\cite{sanchez_detection_2012}, as recently verified~\cite{chida_coulomb_2025}. Time traces of the charge fluctuations reveal the connection of efficiency and crosscorrelations between transported particles and heat quanta transferred from the heat source~\cite{sanchez:2013}. It also demonstrates the purely stochastic nature of the dynamics~\cite{mayrhofer_stochastic_2021}, which does not require any moving or internally oscillating parts. 

When the system-reservoir couplings are larger, higher-order transitions involving electrons in the two dots where the interacting state $(1,1)$ is only virtually occupied~\cite{dare:2017,walldorf:2017,jong_parametric_2021} become important. In this regime, the energy filtering introduced by the discreteness of the quantum dot levels becomes broader, which can help to improve the amount of generated power~\cite{dare:2017}. Since electrons can now tunnel at different energies around those of the quantum dot states, the heat currents are not any longer determined by the particle currents, as in Eq.~\eqref{eq:meJ}. Hence, the correlation between particle and heat currents are reduced and so is the efficiency.

\subsubsection{Quantum correlations}\label{eq:qcorrel}

A few works go further this scheme by considering many body quantum correlations. At very low temperatures, strongly interacting quantum dots exhibit the Kondo effect, characterized by a sharp resonance close to the electrochemical potential due to the screening of the spin of the quantum dot electron by a cloud of opposite spin electrons in the leads~\cite{goldhaberGordon_kondo_1998}. The impact of these in three-terminal energy harvesters (as those discussed in Sec.~\ref{sec:resonant}) has been considered. If the interaction only affects electrons in the same quantum dot, the Kondo peaks introduce additional features to the performance characteristics which do not seem to improve neither the efficiency nor the generated power~\cite{eckern_two_2020}. 
The interaction between the two dots adds a parallel channel for heat which may affect the sign of the individual dot Seebeck coefficients expected for some gate voltage configurations, hence limiting the performance of the system as a thermocouple~\cite{donsa:2014}.

One can also use even-parity superpositions of the occupation of a quantum dot in the proximity of a superconductor to leverage a nonlocal thermoelectric response in the large gap regime, $\Delta\gg\kBT_i$. Consider for instance a configuration of coupled quantum dots as sketched in Fig.~\ref{fig:interact}(b) where terminal $R$ is replaced by a large gap superconductor.   In this regime, the Coulomb interaction interplays with Andreev reflection processes~\cite{martinRodero_josephson_2011}. 
The occupation $n_g=0,1$ of a capacitively coupled (hot) quantum dot, labelled as $g$, affects not only the energy of the even parity system states $E_{\pm,n_g}=A_\mp+\varepsilon_g$, but also the superposition itself: $\left|\pm,n_g\right\rangle \propto E_\pm\left|0,n_g\right\rangle -\Gamma_{S}\left|2,n_g\right\rangle$, where $|n_s,n_g\rangle$ are the dot occupation states, $A_{\pm,n_g}=\tilde\varepsilon_s\pm\sqrt{(\tilde\varepsilon_s+n_gU)^2+\Gamma_S^2}$, $\tilde\varepsilon_s=\varepsilon_s+U/2$, $\varepsilon_\alpha$ is the energy of dot $\alpha\in\{s,g\}$, and $\Gamma_S$ is its coupling to the superconductor. 
Furthermore, fluctuations in $n_g$ induce transitions between them, e.g., $|+,0\rangle\rightarrow|-,1\rangle$~\cite{tabatabaei_nonlocal_2022}. These properties are sufficient to make the contribution of electrons and holes different in the tunneling evens between the system dot and a normal terminal. As a consequence, a finite current is generated between the normal and the superconducting terminals. 
Differently from all normal analogues~\cite{hotspots,holger}, this engine achieves a heat to power conversion without requiring energy dependent tunneling barriers, but rather relying on the stochastic modulation of the system wave function. Additionally, the dependence of $|\pm,n_g\rangle$ on $n_g$ introduces the possibility to modulate the generated current and even its sign by gating the coupled dot.
The same effect can also be used as a refrigerator and for hybrid operations~\cite{lopez_optimal_2023}. 

In the opposite regime, where the gap of the superconductor is comparable to the temperature, one can make use of the density of states as a filter for quasiparticle tunneling~\cite{bhandari:2018,tabatabaei_nonlocal_2022}, a method used for Peltier cooling in normal-insulator-superconductor tunnel junctions~\cite{giazotto:2006}. Remarkably, the current generated this way has an opposite sign to the contribution due to the Andreev-Coulomb contribution discussed above, allowing for probing the dominance of Cooper pairs or of quasiparticles in the transport properties of the conductor~\cite{tabatabaei_nonlocal_2022}. 
Thermoelectric effects in multiterminal hybrid semiconductor-superconductor junctions are reviewed somewhere else~\cite{arrachea_thermoelectric_2025}.

Finally, the nonlocal thermoelectric effect can be used to probe the impact of interactions in the quantum Hall regime. For this, one considers a device with two edge states that are copropagating along a certain distance. With a proper design of quantum point contacts, one can inject electrons from a hot terminal into the inner edge and then separate the two edges such that only the outer edge scatters with a thermoelectric element (e.g., another quantum point contact). If the outer edge connects two terminals at the same temperature, the current between them will be finite only provided some heat has been transferred between the edge channels mediated by the electron-electron interaction of electrons in different channels~\cite{braggio_nonlocal_2024}. Furthermore, one can selectively replace one of the edges at a given point in the interacting region by an equivalent copy, this way avoiding any previously developed crosscorrelations. The generated response is sensitive to this replacement~\cite{braggio_nonlocal_2024}, hence the thermoelectric effect allows for the detection of nonequilibrium crosscorrelations of interacting Luttinger liquids. Recent experiments confirm this picture~\cite{yamazaki_efficient_2025}.

\subsubsection{Information engines and autonomous Maxwell demons}\label{subsec:info}
Interestingly the correlated dynamics of the two coupled subsystems can be interpreted in terms of mutual information~\cite{horowitz:2014prx,koski:2014prl,kutvonen:2016,ptaszynski:2018}, triggering connections with information-driven engines~\cite{whitney_illusory_2023,deOliveiraJunior_friendly_2025}. Indeed, focusing on the capacitively coupled quantum dot systems discussed above~\cite{hotspots}, the two transitions in the source dot, $g$, perform the detection (of the charge in $s$), feedback (the charge in $g$ affecting the tunneling rates in the conductor) and information erasure (restoring the initial state) mechanisms of an autonomous Maxwell demon. This is most clearly evidenced in the equivalent case where the nonlocal thermoelectric response is due to coupling to a colder source~\cite{strasberg:2013}, where the sequence is reversed: $(n_1,n_2)=(0,0)\rightarrow(0,1)\rightarrow(1,1)\rightarrow(1,0)\rightarrow(0,0)$. The cold dot, $g$, is occupied until an electron enters $s$, when it absorbs the energy $U$ to overcome the electrochemical potential $\mu_g$ and tunnel out (detection). This changes the rates in the system (backaction) allowing the electron to tunnel out and restores the initial state (reset). In this case, $U$ is transferred from $s$ to $g$, increasing the entropy of the gate.

Differently from what one expects from a Maxwell demon though, a three terminal configuration cannot reduce the entropy of the conductor by, e.g., generating power, without changing its energy: a finite $J_H\neq0$ is necessarily exchanged via the electron-electron interaction in the quantum dots, see Eq.~\eqref{eq:currhotspot}. References \cite{holger} and \cite{koski:2015} are then rather interpreted as autonomous information-enabled (demonic) heat engines~\cite{whitney_illusory_2023}.  In order to have a conductor subsystem (formed by terminals $L$ and $R$) where the first law is respected while the second is violated, a four terminal configuration is required, with $s$ coupled to two different dots, one hot, the other cold~\cite{whitney:2016,sanchez:2019,fu_quantum_2021,gao_cooling_2023}. 
The thermodynamic discussion is analogous to the one described in Sec.~\ref{subsec:neq} for nonequilibrium states: the entropy reduction in the conductor is compensated by the heat flowing in the coupled terminals. Indeed, if one forces the system quantum dot to reach a thermal state, the demonic effect disappears~\cite{whitney:2016}. However, the role of information is in this case explicit~\cite{sanchez:2019,monsel_autonomous_2025}.

\subsection{Bosonic excitations}\label{sec:ephon}

The case where the injected heat comes from a bosonic environment requires a conductor with the same broken symmetries examined in Sec.~\ref{subsec:ee}, as discussed for a minimal configuration consisting on a single quantum dot (or molecule) coupled to a phonon bath in Ref.~\cite{entin:2010}. Most of the subsequent proposals focus in the hopping regime, where the nanostructure is described by a tight binding approach as a series of localized states. The electrons hop from one to another in their way through the conductor. 
In the case that the nanostructure contains a barrier or a bottleneck transition (e.g., if a site in a one-dimensional chain has a higher energy), transport is suppressed unless additional energy is provided to the electrons from the environment. This environment, that can be treated as a (bosonic) third terminal of the conductor, can take the form of lattice phonons, photons from the electromagnetic fluctuations in nearby conductors or magnons in magnetic insulators, depending on the specific configuration~\cite{girvin_modern_2019}. 

A simple case in which the thermoelectric-generating asymmetries are maximized is a double quantum dot~\cite{brandes_coherent_2005}, with one state below the electrochemical potential ($\varepsilon_L<\mu_L)$ and the other one above ($\varepsilon_R>\mu_R$), as sketched in Fig.~\ref{fig:interact}(c). 
Ignoring the spin of the electron, the system is described by the Hamiltonian
$\hat{H}_{\rm DQD}=\varepsilon_L\hat{d}_L^\dagger\hat{d}_L^{}+\varepsilon_R\hat{d}_R^\dagger\hat{d}_R^{}+\tau(\hat{d}_L^\dagger\hat{d}_R^{}+\hat{d}_R^\dagger\hat{d}_L^{})$, where $\tau$ is the interdot coupling.
When $\varepsilon_R-\varepsilon_L\gg\tau$, the interdot transition is a bottleneck: an electron occupying the left quantum dot needs to overcome the energy difference to hop to the right one and then be transported to terminal $R$. The electron-phonon coupling is given by $\hat{H}_{eph}=\sum_v\lambda_\nu(\hat{n}_L-\hat{n}_R)(\hat{a}_\nu^\dagger+\hat a_{-\nu}^{})$, where $\hat{n}_\alpha=\hat{d}_\alpha^\dagger\hat{d}_\alpha^{}$ is the occupation of dot $\alpha$ and $\hat{a}_\nu^\dagger$ creates a phonon of frequency $\omega_\nu$~\cite{wingreen_inelastic_1989,brandes_spontaneous_1999}. It unblocks the transition via the absorption of a phonon of frequency $\hbar\omega_p\approx\varepsilon_R-\varepsilon_L$ from the phonon bath at temperature $T_p$ and distribution $n_{\rm B}(\omega)=[\exp(\hbar\omega/\kBT_p)-1]^{-1}$. Phonons and electrons are typically at different temperatures in nonequilibrium environments, therefore $T_p-T$ breaks the detail balance of the phonon-assisted tunneling transition. This mechanism enables the direct electric current, which can be controlled by gating the quantum dot levels without requiring energy-dependent tunneling barriers. At $\varepsilon_L=\varepsilon_R$, the system is symmetric and $I=0$. Nonlocal thermoelectric converters have been proposed based on this effect in a variety of setups, including molecules~\cite{entin:2010,entin_three_2012,jiang:2012,jiang:2013prb}, quantum dots~\cite{jiang:2012,goldozian_quantifying_2019} or suspended wires~\cite{jiang:2013prb,bosisio_using_2015,bosisio_nanowire_2016,bosisio_thermoelectric_2017}. In multiterminal (two electronic, two phononic) configurations, interesting connections between phonon drag and pumping have been suggested~\cite{lu_electron_2016}.

Most of these works assume sequential transitions where the electronic phase coherence is lost after every hop. In this regime, it has been shown that the thermoelectric response is defined by the sites coupling to the electronic terminals~\cite{jiang:2013prb}, which determine how much heat is dissipated in the conductor. This is not expected in a phase coherent conductor where the inner sites affect the molecular superpositions~\cite{brandes_coherent_2005}. Respecting the phase coherence at the same time preserving the thermodynamic laws requires a more delicate treatment~\cite{goldozian_quantifying_2019} which may be untreatable as the system size increases. 

This effect has been recently measured in two experiments using a double quantum dot defined in an InAs nanowire~\cite{dorsch:2020,dorsch_characterization_2021}. By tuning the energy levels with plunger gates the current is measured around triple points of the charge stability diagram~\cite{vanderWiel_electron_2002} i.e., where the three states $|n_L,n_R\rangle=\{|0,0\rangle,|1,0\rangle,|0,1\rangle\}$ coexist, as depicted in Fig.~\ref{fig:interact}(c). Local temperature differences can be induced by the dissipated Joule heating from currents along an array of subjacent nanowires~\cite{dorsch_characterization_2021}. A longitudinal temperature difference between the two reservoirs results in a thermoelectric current around the condition $\varepsilon_L\approx\varepsilon_R$ that changes sign around the triple point, as expected for a two terminal conductor~\cite{thierschmann_diffusion_2013}. Additionally, another contribution is measured that changes sign at the triple point in the perpendicular direction: when $\varepsilon-\mu\approx\mu-\varepsilon$ and appears even in the case $T_L=T_R$. This signal is attributed to the phonons being at a different temperature~\cite{dorsch:2020}, which can be confirmed by locally heating at the interdot barrier and measuring a $\unit[]{pA}$ current at temperatures around $\unit[1]{K}$~\cite{dorsch_characterization_2021}. The transverse current (due to heat injected from the environment) can indeed be used to estimate the phonon temperature: the current is suppressed when $|\varepsilon_R-\varepsilon_L|>\kBT_p$~\cite{dorsch:2020}.

Similar effects can be due to coupling to fluctuations of the electromagnetic environment (photons)~\cite{ruokola:2012,henriet_electrical_2015,hofer:2016} or to thermal photons~\cite{rutten:2009,jacob_thermodynamics_2025}. This process can be optimized by coupling a double quantum dot to photons in a cavity~\cite{bergenfeldt:2014,lu_quantum_2019}. The cavity mediated coupling has the practical advantage that the heat bath can be spatially separated from the harvester device over macroscopic distances~\cite{partanen_quantum_2016}, therefore allowing for larger temperature differences between the thermal bath and the conductor. Photons are injected into the cavity by charge fluctuations of an additional double quantum dot coupled to hotter reservoirs. The photon frequency is controlled by the interdot barrier of the hot double quantum dot~\cite{bergenfeldt:2014}. Recent experiments along these lines have measured generated powers of $\unit[1]{fW}$ with 12\% efficiency at electronic temperatures around $\unit[40]{mK}$~\cite{khan_efficient_2021,haldar_microwave_2024}.

Interesting connections can be established with recent proposals of quantum heat engines powered by continuous monitoring~\cite{ferreira_transport_2024,elouard_revealing_2025,sanchez_making_2025}, which renders thermodynamic interpretations of the quantum measurement process~\cite{manzano_quantum_2022,andrew_book}.

\section{Heat currents and cooling}\label{sec:cooling}

The central part of this review has been devoted to nanoscale configurations able to generate electrical power in an isothermal conductor using some resource from a third one (usually heat). However, the same working principles have been proposed to operate these systems for a variety of different useful tasks. These include the nonlocal cooling of a cold reservoir, either using thermal (absorption refrigerators) or voltage (Peltier refrigerators) biases, and thermal management operations aimed at achieving control over heat flows~\cite{jussiau_thermal_2021}. We dedicate this last section to overview recent developments in nonlocal absorption and Peltier refrigerators and thermal management operation exclusive of multiterminal configurations: the circulation of heat and heat transistors.

\subsection{Absorption refrigerators}\label{subsec:absorpt}

A multiterminal nanoscale system can work as an absorption refrigeration when there are at least two different temperatures among the electron reservoirs, so that under appropriate conditions the heat injected by the hottest terminal, $J_H>0$, forces the extraction of heat from the coldest one, $J_C>0$, in the absence of any electrochemical potential bias. In a three-terminal configuration with three temperatures, $T_H>T_0>T_C$, energy conservation ($J_H+J_0+J_C=0)$ and the second law of thermodynamics ($-J_H/T_H-J_0/T_0-J_C/T_C \geq 0)$ impose that heat is dissipated in the reservoir with intermediate temperature, so we can identify $J_H$ as the only resource and define a coefficient of performance (COP) as: $\text{COP}=J_C/J_H$ , which is bounded from above by $\text{COP}_C=T_C/(T_0-T_C)(1-T_0/T_H)$. 

This operation can be implemented in noninteracting resonant tunneling systems as the ones discussed to work as heat engines in Section \ref{sec:resonant}, with the central cavity playing the role of the hot reservoir, and injecting only heat, and the left and right terminals at different temperatures but equal electrochemical potentials. 
If the resonances of the quantum dots are tuned to be both below/above the electrochemical potential, the one of the hot cavity will increase/decrease to fulfill particle conservation, this way extracting heat from the cold reservoir~\cite{manikandan_autonomous_2020}.

\begin{figure*}[t]
\centering
\includegraphics[width=.85\textwidth]{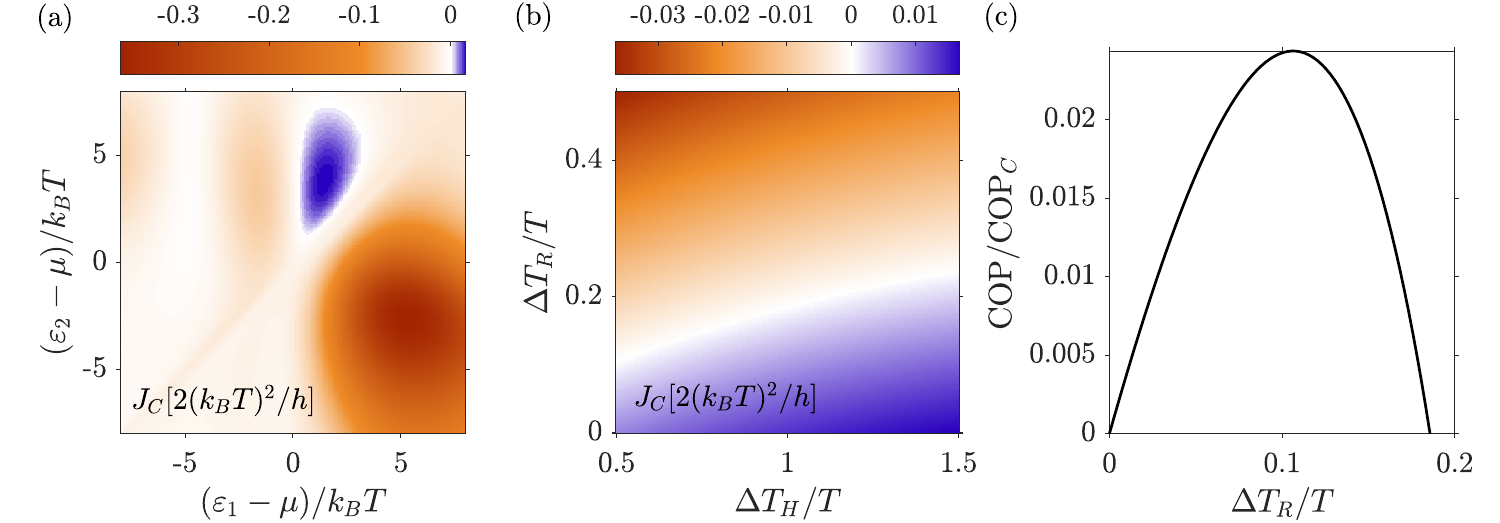}
\caption{Phase coherent absorption refrigerator. Heat is extracted from the cold reservoir of a three terminal system as the one depicted in Figure \ref{fig:elast}(a), with scattering regions 1 and 2 being quantum dots with resonance energies $\varepsilon_1$ and $\varepsilon_2$ and broadening $\Gamma=0.2k_BT$. (a) Cooling power, $J_C=J_L$, as a function of the resonance energies produced by a hot tip with $\Delta T_H/T=1$, and isothermal terminals $L$ and $R$, $T_L=T_R=T$. (b) Cooling power as a function of the heating at the tip, $\Delta T_H$, and of the right terminal, $\Delta T_R$, for $T_L=T$ and $(\varepsilon_1-\mu)/k_BT=2$, $(\varepsilon_2-\mu)/k_BT=4$. (c) Coefficient of performance for fixed $\Delta T_H=1$ but otherwise same parameters as in (b). Fixed parameters: $\mu=20 k_BT$, tip transmission $\epsilon=0.5$, $d=2.5l_0$, $x=1.7l_0$, being $l_0=\hbar/\sqrt{8m\kBT}$, with the electron mass $m$.
}\label{fig:absrefr}
\end{figure*}

So far, Ref.~\cite{manikandan_autonomous_2020} is the only absorption refrigerator in a noninteracting conductor to the best to our knowledge. The reason may be that this effect is very sensitive to the elastic coupling between terminals $L$ and $R$. We verified for this review that the same mechanism works for the fully phase-coherent implementation of the noninteracting resonant tunneling thermocouple discussed in Sec.~\ref{subsec:3t} and sketched in Fig.~\ref{fig:elast}(a).
The tip is used as a heat source that refrigerates one of the conductor terminals depending on its position. In Fig.~\ref{fig:absrefr}(a) we show the heat flow $J_L$ when the tip temperature is increased by $\Delta T_H$ while maintaining $T_L=T_R=T$, as a function of the resonant level energies of the quantum dots. 
Cooling occurs for a very specific configuration where both levels are above the electrochemical potential, $\varepsilon_\alpha>\mu$, and destructive interference suppresses ${\cal T}_{LR}(E)$. In contrast to the inelastic version, no refrigeration is found in the case with $\varepsilon_\alpha<\mu$ (note that symmetries of the transmission probabilities exchanging $\varepsilon_1\leftrightarrow\varepsilon_2$ are only expected upon moving the tip from $x$ to $d-x$). The effect is also sensitive to an increase of the temperature of the other terminal, $\Delta T_R$, as shown in Fig.~\ref{fig:absrefr}(b), as expected. The coefficient of performance is shown in Fig.~\ref{fig:absrefr}(c) as a function of this temperature, attaining only a low fraction of the Carnot limit for this non optimized choice of parameters.

A more robust effect is found due to interactions. Capacitively coupled quantum dots in the arrangement of Fig. \ref{fig:interact}(a) work as absorption refrigerators when the reservoir coupled to quantum dot $g$ is either the heat injector~\cite{hotspots} or the refrigerated cold terminal~\cite{dare:2019,erdman_absorption_2018}. In the former case, it is sufficient that the heat conversion engine discussed in Sec.~\ref{subsec:ee} operates in a configuration in which the generated current either extracts electrons from a reservoir ($L$ or $R$) above its electrochemical potential, or injects them below it~\cite{hotspots}. Interestingly, the refrigeration of one conductor terminal can be combined with pumping heat on the hotter (but not hotter than the gate terminal) and additionally generating power, such that all currents in the conductor are reversed~\cite{manzano_hybrid_2020}. 
In the later case (cooling the terminal coupled to $g$), their operation relies in using the hot temperature of one of the terminals coupled to $s$ to favor the same series of fluctuations, $(n_s,n_g)=(0,0)\rightarrow(1,0)\rightarrow(1,1)\rightarrow(0,1)\rightarrow(0,0)$, cf. Eq.~\eqref{eq:sequence}, that was used to transfer an energy quantum $U$ from $g$ to $s$ to power the heat engine. However, this requires a non trivial choice of the system parameters that limits its operating regime, and is restricted to low cooling powers.

A photonic environment can also play the role of the hot resource for absorption refrigerators. Thermal photons coupled to a double quantum dot connecting two electronic reservoirs at different temperatures can drive processes that extract heat for the colder one \cite{cleuren_cooling_2012,lu_brownian_2020}. The inverse scenario, cooling a photonic resonator using two electronic reservoirs at different temperatures, was also explored in hybrid (superconductor-metal) configurations connected by a quantum dot \cite{tabatabaei_quantum_2024}.

\subsection{Nonlocal Peltier cooling}\label{subsec:nonolcPeltier}

Energy harvesting and nonlocal Peltier refrigeration are related by the Onsager reciprocity relations. The same process that converts heat into power can be reversed to extract heat by powering a conductor. In the former case, heat injected from the third terminal generates a current. In the later, driving current in the conductor extracts heat from the third bath. Only the driving force changes: A temperature difference or a chemical potential bias, respectively.  
Thus, in this case, the resource is the consumed power, so the typical coefficient of performance is $\text{COP}=-J_{_i}/P$, where $J_i>0$ is the heat extracted from terminal $i$. 

To show this more clearly, it is once more useful to consider the linear response regime. Neglecting direct transport between $L$ and $R$, the heat current in the third reservoir (let us call it $C$, for cold, in this case) is
\begin{equation}
J_C=\frac{M_{CL}G_{CR}-M_{CR}G_{CL}}{G_{CL}+G_{CR}}(\mu_L-\mu_R).
\end{equation}
Non surprisingly, in the absence of a magnetic field, replacing $M_{ij}=TL_{ij}$, we can relate the above $J_C/(\mu_L-\mu_R)$ with the ratio $I_L/\Delta T_H$ in Eq.~\eqref{eq:inelasticlinear}.

This behaviour was first predicted~\cite{edwards:1993,edwards_cryogenic_1995} and experimentally realized~\cite{prance:2009} for cooling a small but microscopic region of a 2DEG sandwiched by resonant tunneling barriers (in the same disposition of the noninteracting thermocouples discussed in Sections \ref{sec:resonant} and \ref{subsec:absorpt}). The central cavity is cooled when it is connected to the terminal with higher/lower electrochemical potential (but same temperature) by a quantum dot whose resonance is below/above its own electrochemical potential. This way a particle current is generated in favor of the voltage bias (dissipating power in $L$ and $R$), supported by cold electrons entering the central region and hot ones leaving it, yielding a decrease in its temperature. A reduction of $\unit[90]{mK}$ at base temperatures of around $\unit[120]{mK}$ was achieved~\cite{prance:2009}. 

The gap and quasiparticle resonances can be used in normal-superconductor interfaces~\cite{Nahum1994Dec,Leivo1996Apr}.
Here, phase-coherent three-terminal configurations have been suggested to enhance the cooling power in a normal terminal coupled to two superconductors via destructive Andreev interference~\cite{cioni_high_2025}: A phase difference $\pi$ between the superconductors suppresses the Joule heat contribution of subgap currents.

Analogously, the refrigeration side of chiral and Aharonov-Bohm engines discussed in Secs.~\ref{subsec:chiral} and \ref{subsec:3t}, has been discussed using the split resonances of two quantum dots and taking advantage of the directionality imposed by the magnetic field~\cite{sanchez_nonlinear_2019,jimenezValencia_persistent_2025}.

Capacitively coupled quantum dot configurations, as those discussed in Sections \ref{subsec:ee} and \ref{subsec:absorpt}, have also been explored as Peltier refrigerators \cite{hotspots,zhang:2015,hussein_heat_2016,walldorf:2017,bhandari:2018,dare:2019,tabatabaei_nonlocal_2022}. 
Again one uses the dynamic connection between particles tunneling in $s$ and energy quanta $U$ exchanged with $g$ established by the tunneling asymmetries in the conductor. 
In this case the difference of electrochemical potential $\mu_L-\mu_R$ is used to favor series of fluctuations that are able to reverse the heat flow. 
Depending on the specific tuning of the quantum dot resonances, this can be used to cool one of the system reservoirs. The electrochemical potential bias is found to be much more efficient and robust than the temperature difference used in absorption refrigerators to favor a suitable series of fluctuations \cite{dare:2019}. 
The effect of coherent cotunneling processes on the performance of capacitively coupled quantum dots Peltier refrigerators has been found to give an overall reduction in cooling power as these processes do not share the energy selectivity of the sequential tunneling~\cite{walldorf:2017}. 

A somewhat similar configuration was realized experimentally using capacitively coupled metallic islands with energy-independent (i.e., exhibiting no thermoelectric effect) tunneling barriers~\cite{koski:2015}. In that case, cooling $L$ and $R$ by about $\unit[1]{mK}$ at temperatures of $\unit[77]{mK}$ is possible only if the gate terminal is cold~\cite{thierschmann_thermoelectrics_2016}. This configuration is therefore not a Peltier refrigerator: there is no thermoelectric effect involved (not Peltier) and heat is not extracted from the coldest reservoir (not a refrigerator). The remarkable feature is that the capacitive coupling is able to remove the Joule heat from the particle conductor (where power would be otherwise dissipated). This process was interpreted in terms of autonomous Maxwell demon operations~\cite{koski:2015}, along the lines discussed in Section \ref{subsec:info}. 

In metallic islands, the energy balance needs to take the phonon bath into account, which adds an additional reservoir to the problem. Using the quasiparticle spectral properties of superconductor terminals in the current-carrying conductor, the temperature of a capacitively coupled single-electron box has been predicted to be reduced from $\unit[200]{mK}$ to $\unit[50]{mK}$ for islands of size $\unit[30\times50\times300]{nm^3}$~\cite{sanchez_correlation_2017}. 
One can also extract heat from the phonon bath that couples to the nanostructure~\cite{lu_unconventional_2021,lu_multitask_2023}, with similar ideas used for cooling of molecular vibrations close to their ground state~\cite{zippilli_cooling_2009,santandrea_cooling_2011,simine_vibrational_2012,stadler_ground_2014,arrachea_vibrational_2014,stadler_ground_2016,mukherjee_three_2020,sevitz_autonomous_2025}, as demonstrated in nanotube resonators~\cite{urgell_cooling_2020}.

\subsection{Thermal management}
\label{subsec:manage}

In this section we discuss the propagation of heat and how to manage it with nanoscale three terminal conductors by modulating its direction (circulators) or its amplitude (transistors).

\subsubsection{Heat circulators}
\label{subsec:circul}

The three terminal configuration also enables the possibility of defining circulation operations, where the heat injected from one terminal flows preferably in one direction (clock- or counterclockwise) to the next one, while transport in the opposite way is partially or totally blocked. A clear case where this happens is in chiral conductors under strong magnetic fields: with the heat through the bulk of a quantum Hall bar being suppressed~\footnote{Limitations of this picture have been discussed~\cite{marguerite_imaging_2019,melcer_heat_2024}.}, the absence of backscattering leads the heat to be transferred only between chirally connected reservoirs~\cite{granger_observation_2009,chiraldiode}, as illustrated in Fig.~\ref{fig:chiral}(a).

For smaller magnetic fields, this behavior arises due to Aharonov-Bohm interference in ring configurations, both in continuum \cite{adrian} or tight binding models with normal or superconducting leads~\cite{acciai:2021, hwang:2018}. This is because the broken time-reversal symmetry naturally imposes a preferred transmission, i.e., in $\mathcal{T}_{jH}(B)\neq\mathcal{T}_{Hj}(B)=\mathcal{T}_{jH}(-B)$, with $j=L,R$.
In the presence of a magnetic field, electrons acquire a geometric phase that can only play a role in a coherent system if they go around a closed loop via the Aharonov-Bohm effect. 
The interplay of the geometric and kinetic phases can be modulated with the flux piercing the ring and with gate voltages, allowing to find conditions with destructive or constructive interference for selected transmissions. This way, circulation coefficients close to unity can be achieved~\cite{acciai:2021,adrian}. The system can be operated in an all-thermal way, with no particle current flowing in the system on average, with a slightly smaller efficiency~\cite{adrian}. 

As this mechanism relies on the phase coherence of the electron waves, it is expected that the effect is reduced with the presence of dephasing mechanisms~\cite{adrian}. 
The addition of superconducting terminals confers a higher system tunability to the devices via the possibility to establish superconducting phase biases~\cite{acciai:2021}, but they do not seem to improve the performance of normal configurations.

Other operations related to heat rectification in which the current between two conductors is not reciprocal under swapping their temperatures have been proposed that rely on heat dissipated in the additional reservoirs~\cite{martinezPerez_rectification_2015,jiang:2015,guillem,zhang_three_2017,sanchez_single_2017,genevieve,extrinsic,tesser_heat_2022,wysokinski_four_2023}.

\subsubsection{Thermal transistors}
\label{susec:transistors}

A typical three-terminal device is the transistor: a current between two terminals is modulated by tuning the third terminal which acts the base. A thermal transistor is able to modulate a heat current by tuning the base temperature~\cite{li_negative_2006}. Ideally, small variations in this temperature result in large transport modulations. 

Thermal gating of a particle current was observed in a system of capacitively coupled quantum dots~\cite{thierschmann_thermal_2015}, see Fig.~\ref{fig:interact}(a) for an illustration. For low bias $|\Delta\mu|<U$, fluctuations in the occupation of the gate dot (which strongly depend on the temperature of its terminal) take the system dot in and out of the transport window, this way affecting the current by $\unit[\pm30]{pA}$~\cite{thierschmann_thermoelectrics_2016}. The larger the temperature, the larger the modulation. 

In order to avoid that the thermal gating involves a large transfer of heat, one needs to filter the conductor transitions~\cite{sanchez_all-thermal_2017,sanchez_single_2017}. Consider the configuration described in Sec.~\ref{subsec:ee}, where the tunneling rates in the quantum dot $s$, $\Gamma_{n_g}^i$ ($i=L,R$), depend on the occupation, $n_g$, of quantum dot $g$. In the case that they are suppressed for a particular occupation of the dot (e.g., $\Gamma_1^i\ll\Gamma_0^i$), a fluctuation in the gate dot occupation switches on/off the transport through $s$. Furthermore, electrons entering $s$ will leave it with the same energy ($\varepsilon_s$, in this case): i.e., the current modulation does not require any heat transfer from the gate~\cite{sanchez_all-thermal_2017}. The necessary filtering can be done either with additional quantum dots~\cite{sanchez_single_2017} or using the spectral properties of the reservoirs. Using the gap of two superconducting terminals, the combination of two such thermal transistors at different temperatures was shown to enable a Maxwell demon~\cite{sanchez:2019}. 
These systems benefit from the energy exchange being quantized in units of the electron-electron interaction, $U$. 

An alternative is to consider the sensitivity of a quantum point contact to the charge of a nearby quantum dot~\cite{vandersypen:2004} (the base): near pinch off, charging the base brings an open channel into the tunneling regime~\cite{yang_thermal_2019}.
Bosonic base terminals have also been suggested~\cite{jiang:2015,lu_quantum_2019,lu_brownian_2020} which require more complicated settings to achieve the proper filtering.

\section{Discussion}\label{sec:conclusion}
The development of nanoscale devices with designed spectral properties gives access to the dynamics of few and even single particles and to manipulate them depending on their energy. This introduces mechanisms to control heat flows in quantum systems and use them for thermodynamic operations such as low power quantum heat engines, refrigerators or heat management devices. While their practical use in macroscopic setups is compromised by scalability and restricted to low temperatures, they could be incorporated to work onchip in quantum technologies, assisting on their energetics~\cite{auffeves_quantum_2022}. For this, multiterminal configurations able to separate the particle and heat flows are convenient versions of the thermocouple. We have reviewed different configurations of nanoscale energy harvesters, absorption and Peltier refrigerators and thermal circulators based in assorted configurations exploiting phase coherence, chirality, single-particle interactions an nonthermal states. Thermoelectric effects also allow to investigate properties that are opaque to conductance measurements such as the role of correlations or of mutual information, having hence and important role in fundamental research.

\acknowledgements

We thank J.-H. Jiang, J. Lu and B. Sothmann for useful comments on the manuscript, and acknowledge collaborations over the years which led to some of the works discussed here, involving M. Acciai, B. Ben\'itez, A. Braggio, P. Burset, M. Carrega, C. Elouard, G. Fleury, C. Gorini, D. Goury, G. Haack, F. Hajiloo, F. Haupt, G. Jaliel, A. N. Jordan, A. Levy Yeyati, R. L\'opez, G. Manzano, A. Mecha, L. W. Molenkamp, J. Monsel, P. P. Potts, R. K. Puddy, P. Samuelsson, J. Splettstoesser, D. S\'anchez, C. G. Smith,  B. Sothmann, S. M. Tabatabaei, L. Tesser, H. Thierschmann, R. S. Whitney. J. Yang, and the late M. B\"uttiker, with whom it was started.  We acknowledge funding from the Spanish Ministerio de Ciencia e Innovaci\'on via grants No. PID2022-142911NB-I00 and No. PID2024-157821NB-I0, and through the ``Mar\'{i}a de Maeztu'' Programme for Units of Excellence in R{\&}D CEX2023-001316-M.





\bibliography{biblio}

\end{document}